\documentclass{nature}

\usepackage{graphicx}
\usepackage{amssymb}
\usepackage{hyperref}

\title{Terrestrial planet formation by torque-driven\\[8pt] convergent migration of planetary embryos}

\author{M.~Bro\v z$^1$, O.~Chrenko$^1$, D.~Nesvorn\'y$^2$, N. Dauphas$^3$}

\def\d{{\rm d}}

\begin{document}

\let\includegraphics\oldincludegraphics

\maketitle

\begin{affiliations}
 \item Institute of Astronomy, Charles University, V Hole\v sovi\v ck\'ach 2, 18000 Prague 8, Czech Republic
 \item Department of Space Studies, Southwest Research Institute, 1050 Walnut St., Suite 300, Boulder, CO 80302, USA
 \item Origins Laboratory, Department of the Geophysical Sciences and Enrico Fermi Institute, The University of Chicago, 5734 South Ellis Av., Chicago, IL 60637, USA

\end{affiliations}

%% 1st paragraph (Abstract)

\begin{abstract}
Massive cores of the giant planets are thought to have formed in a gas disk by accretion
of pebble-size particles whose accretional cross-section is enhanced
by aerodynamic gas drag\cite{Lambrechts_Johansen_2012A&A...544A..32L,Levison_etal_2015Natur.524..322L}.
A commonly held view is that the terrestrial planet system formed later
(30--200\,Myr after the dispersal of the gas disk) by giant collisions
of tens of roughly Mars-size protoplanets\cite{Chambers_Wetherill_1998Icar..136..304C}.
Here we propose, instead, that the terrestrial planets formed earlier
by gas-driven convergent migration of protoplanets toward $\sim\!1\,{\rm au}$
(related ref.~\cite{Ogihara_etal_2018A&A...612L...5O} invoked a different process to concentrate planetesimals).
To investigate situations in which convergent migration occurs, 
we developed a~radiation-hydrodynamic model with realistic opacities\cite{Zhu_etal_2012ApJ...746..110Z,Semenov_etal_2003A&A...410..611S}
to determine the thermal structure of the gas and pebble disks
in the terrestrial planet zone. We find that protoplanets rapidly grow 
by mutual collisions and pebble accretion, and gain orbital eccentricities
by gravitational scattering and the hot-trail effect\cite{Chrenko_etal_2017A&A...606A.114C,Eklund_Masset_2017MNRAS.469..206E}.
The orbital structure of the terrestrial planet system is well reproduced
in our simulations, including its tight mass concentration at 0.7--1\,au
and the small sizes of Mercury and Mars. 
The early-stage protosolar disk temperature exceeds 1500\,K inside 0.4\,au
implying that Mercury grew in a highly reducing environment,
next to the evaporation lines of iron and silicates,
influencing Mercury's bulk composition\cite{Hauck_etal_2013JGRE..118.1204H}. 
In a late-stage cold gas disk, accretion of icy/hydrated pebbles
would contribute to Earth's water budget.
\end{abstract}

%% 2nd paragraph (Problem)

In the standard model of the terrestrial planet formation\cite{Wetherill_1990AREPS..18..205W},
accretional collisions between ${\sim}\,1$~to $100{\rm km}$ planetesimals lead to gradual build up of
lunar- to Mars-size protoplanets, which gravitationally interact and further
grow during a {\em late\/} stage of giant impacts\cite{Chambers_Wetherill_1998Icar..136..304C}.
Indeed, radiometric data indicate that the Moon-forming impact on proto-Earth\cite{Canup_etal_2019} probably happened relatively late,
some $50\,{\rm Myr}$ or more after the appearance of the first solar system solids\cite{Thiemens_etal_2019NatGe..12..696T}.
Computer simulations of late-stage accretion are required to match various constraints,
including the similar semimajor axes of Venus and Earth (only ${\simeq}\,0.3\,{\rm au}$ radial separation).
The results indicate that the terrestrial planets should have accreted from
a narrow annulus (0.7--1\,au)\cite{Hansen_2009ApJ...703.1131H},
perhaps because the planetesimal disk was truncated at ${\sim}\,1\,{\rm au}$ by Jupiter's
inward and then outward migration\cite{Walsh_etal_2011Natur.475..206W}.
The cause behind the suggested inner edge at $0.7\,{\rm au}$ is less clear,
but it might be induced by an `inverted' gas surface density profile $\Sigma(r)$
which would initially concentrate planetesimals by gas drag\cite{Ogihara_etal_2018A&A...612L...5O}.
Since a narrow annulus of planetesimals tends to spread over time, however,
it is still difficult for those models to reproduce the tight terrestrial 
planet mass concentration at 0.7--1\,au\cite{Deienno_etal_2019ApJ...876..103D}.
Moreover, the early growth of terrestrial
protoplanets may have been regulated by efficient accretion of small, 
pebble-size particles\cite{Levison_etal_2015PNAS..11214180L,Johansen_etal_2015SciA....1E0109J},
rather than larger planetesimals. Measurements of isotopic
anomalies also indicate that the terrestrial planets may have formed
faster than in the standard model, possibly while the gas disk was still
around\cite{Dauphas_Pourmand_2011Natur.473..489D}.

%% 3rd paragraph (Results)
%% marssize

Gas-driven migration is critically important for protoplanets\cite{Tanaka_Ward_2004ApJ...602..388T}.
We constructed a radiation-hydrodynamic model of gas and pebble disks to 
study migration in the terrestrial planet zone.
In our model, hereafter named {\it Thorin\/}\cite{Chrenko_etal_2017A&A...606A.114C}, 
protoplanets migrate due to several gravitational
torque terms (Lindblad, corotation and heating\cite{Benitez_etal_2015Natur.520...63B}).
Protoplanets can grow in mutual collisions and by accreting pebbles but not gas.
The simulations included Mercury- to Mars-size protoplanets ($0.05\hbox{ to }0.1\,M_{\rm E}$,
where $M_{\rm E}=6\times10^{24}$\,kg is the Earth mass),
initially located in a wide annulus (0.4--1.8\,au; Figure~1).
Our fiducial gas disk model\cite{Kretke_Lin_2012ApJ...755...74K}
is defined by the radial gas flux~$\dot M$ and kinematic viscosity~$\nu(r)$,
which is a function of radius~$r$.
The disk exhibits an increasing surface density $\Sigma(r)$ with~$r$
accross the inner portion of the terrestrial zone.
This choice is motivated by studies of disks
with magneto-rotational instability (MRI)\cite{Flock_etal_2017ApJ...835..230F},
layered accretion\cite{Kretke_Lin_2012ApJ...755...74K}, or
disk winds\cite{Ogihara_etal_2018A&A...612L...5O},
where similar reversal of the $\Sigma(r)$ profile happens.
The gas temperature~$T$ and disk aspect ratio $h = H/r$
are controlled by the opacity $\kappa(\rho,T)$\cite{Zhu_etal_2012ApJ...746..110Z},
where $\rho$ is the gas density. The pebble disk is defined by
the pebble flux~$\dot M_{\rm p}$ at the outer boundary.
All these components affect migration of protoplanets.
See Methods and the supplementary information (SI)
for a complete model description.

The simulations were used to compute the migration speed and direction 
of protoplanets as a~function of their mass $m$ and semimajor axis~$a$
(Figure~2)\cite{Paardekooper_etal_2011MNRAS.410..293P}.
We found that protoplanets with $a>1\,{\rm au}$ migrate inward
and protoplanets with $a<1\,{\rm au}$ migrate outward.
The absolute value of the torque generally scales as
$\Gamma_0 = (q/h)^2\Sigma r^4\Omega^2$,
where $q = m/M_{\rm S}$~is the planet-to-Sun mass ratio and
$\Omega$~the Keplerian angular velocity,
but its precise value and sign depend on details
how gas flows in the vicinity of each protoplanet.
The Lindblad torque is induced by resonant density waves
which form two spiral arms. It is usually negative and leads
to the inward migration, but it may reduce and {\em reverse}
if the eccentricity is excited to $e > 2h$,
e.g., during close encounters between protoplanets.
A semi-analytical expression of \cite{Paardekooper_etal_2011MNRAS.410..293P}
reveals its dependence on the slopes $\alpha$, $\beta$ of $\Sigma$, $T$ profiles,
$\Gamma_{\rm L} = \Gamma_0/\gamma_{\rm diff}\,(-2.5-1.7\beta+0.1\alpha)(0.4h/b_{\rm sm})^{0.71}$;
$b_{\rm sm} = r_{\rm sm}/(hr)$, where
$r_{\rm sm}$ denotes the smoothing length and
$\gamma_{\rm diff}$ the diffusion coefficient.
The corotation torque is related to gas moving along horseshoe-like orbits.
It becomes positive for certain slopes of $\Sigma$, $T$,
if it is kept unsaturated, e.g., by non-zero viscosity.
The corresponding expression\cite{Paardekooper_etal_2011MNRAS.410..293P} is
$\Gamma_{\rm c} = \Gamma_0/\gamma_{\rm diff}[(1.5-\alpha-2\xi/\gamma_{\rm diff})(0.4h/b_{\rm sm})^{1.26} + 2.2\xi(0.4h/b_{\rm sm})^{0.71}]$,
where $\xi \equiv \beta-(\gamma-1)\alpha$ is the slope of entropy~${\cal S}(r)$.
The heating torque\cite{Benitez_etal_2015Natur.520...63B} arises
when the gas flow is heated by an accreting protoplanet,
which creates an underdense region behind it and thus a positive torque contribution.
Taken together, these torques form a convergence zone
in our fiducial disk model.
The convergent migration must lead to the accumulation of planets near
the convergent radius $r_{\rm c} \simeq 1\,{\rm au}$.
The characteristic migration rates computed numerically by our RHD model are
${\rm d}a/{\rm d}t \simeq \pm 10^{-7}\hbox{ to }10^{-6}\,{\rm au}\,{\rm yr}^{-1}$.
The protoplanets are therefore expected to move near $r_{\rm c}$ well within the lifetime of the protosolar disk (3--10\,Myr).
Based on this result we suggest that the convergent migration
of protoplanets is the main reason behind the concentration
of the terrestrial planet mass at 0.7--1\,au.
Planet migration in different disk models (e.g. the Minimum Mass Solar Nebula;
MMSN) is discussed in SI.

%% Symba
%% Mercury eccentricity.
%% Mars eccentricity.

We performed long-term accretional simulations with a symplectic N-body integrator
called {\it SyMBA\/}\cite{Duncan_etal_1998AJ....116.2067D}
to examine the effect of convergent migration during
the gas disk lifetime on the growth of planets.
The radial migration of protoplanets,
including the evolution of eccentricity and inclination\cite{Tanaka_Ward_2004ApJ...602..388T,Chrenko_etal_2017A&A...606A.114C,Eklund_Masset_2017MNRAS.469..206E},
was informed from our hydrocode simulations described above
and mimicked by additional acceleration terms in {\it SyMBA\/} (Methods).
Specifically, the migration rate ${\rm d}a/{\rm d}t$ was set as a function of
planet mass and orbit (from Figure~2 and similar results that we obtained for
other disk models; SI). We found that convergent migration leads to
rapid accretion of planets with properties that often match
the real terrestrial planet system (Figure~3). The results vary, however, due
to the stochastic nature of accretion and we thus performed dozens
of simulations to statistically characterize them.
We found that:
(i)~3~to 6 terrestrial planets typically form in $<$10\,Myr,
(ii)~planet mass is concentrated at ${\simeq}\,$0.7--1\,au,
(iii)~low-mass, Mercury-to-Mars planets end up on moderately excited
orbits at $<$0.7\,au or $>$1\,au, and
(iv)~collisions between similar-sized protoplanets are relatively common (see also~\cite{Canup_etal_2019}).
The system architecture is established by collisions, differential migration and
eccentricity damping\cite{Tanaka_Ward_2004ApJ...602..388T},
and is quite sensitive to the timing of the gas disk dispersal.
The relative importance of pebble accretion for planet growth 
depends on the time-integrated pebble flux. For example, for a constant 
pebble flux $\dot M_{\rm p} = 2\times10^{-6}\,M_{\rm E}\,{\rm yr}^{-1}$,
the growth is roughly equally contributed by pebbles and protoplanet 
mergers. Significantly larger pebble fluxes sustained over millions 
of years would lead to planet overgrowth (i.e., super Earths formation) 
and would weaken radial mass concentration.
A~more systematic sampling of model parameters could help
to `reverse engineer' the disk conditions and lifetime.

At the heart of our argument, the convergent migration of protoplanets 
produced just the right conditions, with a strong mass concentration near 0.7--1\,au,
for the formation of the terrestrial planets\cite{Hansen_2009ApJ...703.1131H}.
This new model offers a notable advantage over the previously suggested
mechanisms of annulus truncation\cite{Walsh_etal_2011Natur.475..206W}, because:
(i)~convergent migration confines the annulus from {\it both} sides, and
(ii)~Venus and Earth remain within the convergence zone,
avoiding problems with the radial mass spreading\cite{Deienno_etal_2019ApJ...876..103D}.

%% 5planet_1over10

A gas disk nearing the end of its lifespan is expected to be rarefied
by viscous spreading/photoevaporation, and consequently cold\cite{Bitsch_etal_2015A&A...575A..28B}.
To study the orbital behavior of the terrestrial planets just before dispersal of the gas,
we evaluated the effect of setting the surface density to values
10 or 100 times lower than the nominal value,
$\Sigma_0 = 750\,{\rm g}\,{\rm cm}^{-2}$ at 1\,au.
The terrestrial planets were assumed to be nearly formed
and close to their present orbital radii.
A pebble accretion flux up to $\dot M_{\rm p} \simeq 2\times10^{-4}\,M_{\rm E}\,{\rm yr}^{-1}$
was adopted in {\it Thorin\/}.
Significantly, we found that the planetary orbits became excited
by the hot-trail effect\cite{Chrenko_etal_2017A&A...606A.114C,Eklund_Masset_2017MNRAS.469..206E},
which arises due to pebble accretion on protoplanets,
release of their kinetic energy, and
radiative heating of neighbour gas (see Figure~4).
It is especially effective for $\sim$10 times lower $\Sigma$,
when the thermal capacity of gas is lower
but its gravity still substantial.
As a result, the orbital eccentricities of planets evolve toward
asymptotic values ${\simeq}\,$0.015--0.02 and migrating planets avoid capture in orbital resonances.
The hot-trail effect can explain the current orbital eccentricities of Venus and Earth
(proper $e=0.02$ and 0.01, respectively),
which was never suggested before.
Additional changes may have been inflicted in the terrestrial planet system
by gravitational perturbations during migration/instability of the giant planets\cite{Clement_etal_2018Icar..311..340C}.

%% Moon origin.

A strict version of this work's main thesis --the terrestrial planets formed early--
would imply that the Moon-forming impact occurred early as well ($t_{\rm Moon}<10\,{\rm Myr}$).
Lunar Magma Ocean (LMO) solidification may have occurred much later,
because the LMO may have been sustained by tidal heating for hundreds of million years\cite{Elkins-Tanton_2012AREPS..40..113E}.
However, for $t_{\rm Moon}<10\,{\rm Myr}$,  
geochemical modeling of the Hf/W system (ref.~\cite{Fischer_Nimmo_2018E&PSL.499..257F,Thiemens_etal_2019NatGe..12..696T} 
and SI) shows that the tungsten anomaly in the mantle of Earth would be generally 
higher than the observed value, $\varepsilon_{\rm 182W} = 1.9\pm0.1$.
We therefore prefer our simulations with {\it SyMBA} that ended with five 
(or more) terrestrial protoplanets, thus leaving space for a late Moon-forming 
impact and $\varepsilon_{\rm 182W}$ decrease through equilibration.
A late impact could have spontaneously occurred in a dynamically unstable 
terrestrial system or been triggered by outer planet migration/instability\cite{Roig_etal_2016ApJ...820L..30R}.

%% Small Mercury.
%% Mercury iron core.

The early formation of the terrestrial planets in a gas disk has several 
important implications. For example, the innermost part of a viscously heated
disk can reach the evaporation threshold, $T_{\rm ev}$,
of many minerals. As solids evaporate at the critical
radius~$r_{\rm ev}$, they cannot contribute to planet's growth
below that radius. This could help to explain the small mass of Mercury.
Temperatures in a massive, early-stage protosolar disk shown in Figure~4
reach $T \simeq 1500\,{\rm K}$ and create a highly reducing environment
in which pebbles drift from larger radii\cite{Johansen_etal_2015SciA....1E0109J}
down to $r_{\rm ev}\sim0.4\,{\rm au}$.
Evaporation of pebbles alters the local chemical composition of the gas.
This is very different from previous nebular hypotheses
which dealt only with a narrow ring of local material.
Together with nebular metal-silicate fractionation,
it could naturally explain the large Fe core of Mercury\cite{Hauck_etal_2013JGRE..118.1204H}
and relax constraints on the hypothesized impact-related removal of the 
silicate mantle\cite{Asphaug_Reufer_2014NatGe...7..564A}.
Alternately, in the impact hypothesis, Mercury cannot re-accrete dispersed silicates,
which could be more easily achieved if the stripping of Mercury's
mantle occurred before dissipation of nebular gas\cite{Nittler_etal_2018}.
As the temperature decreases in a low-mass, late-stage disk (Figure~4), 
Mercury's surface could have been enriched in volatiles (e.g., Na, S, K, Cl;
delivered by pebbles), as needed to explain its large volatile budget\cite{Nittler_etal_2018}.

%% Water delivery.

Towards the end of the disk's lifetime, the gas density and viscous heating
decrease, but the disk midplane is still shadowed, thus allowing the snowline
to move down to ${\sim}\,1\,{\rm au}$\cite{Bitsch_etal_2015A&A...575A..28B}.
This could create the right conditions for water delivery to the Earth
if the flux of icy/hydrated pebbles from ${>}\,3\,{\rm au}$ remained
sufficiently high for sufficiently long time.
For reference, we calculate that the Earth would accrete 1--1.5\% of
pebbles moving past 1\,au. To deliver 1~Earth ocean worth of water
($2.3\times10^{-4}\,M_{\rm E}$), we would need that $f \dot M_{\rm p} \delta t
\sim 0.02\,M_{\rm E}$, where $0<f<1$ is the mass fraction of water
in pebbles, $\dot M_{\rm p}$ is the pebble flux at 1\,au and $\delta t$
is the time interval for which the pebble flux is sustained in a cold disk.
This can be achieved, for example, for $\dot M_{\rm p} \simeq 2\times
10^{-6}\,M_{\rm E}\,{\rm yr}^{-1}$ and $f=1$ in mere $\delta t=10^4\,{\rm yr}$. 
The total water content in the Earth, however, is estimated to be equivalent 
to 2--8~Earth oceans\cite{Peslier_etal_2017SSRv..212..743P}, which would 
imply a proportionally longer timescale or other delivery methods\cite{Walsh_etal_2011Natur.475..206W}.
Also, the importance of this process would diminish if Jupiter and Saturn
block the flux of icy pebbles from
${>}\,5\,{\rm au}$\cite{Morbidelli_etal_2016Icar..267..368M}.

%% Outlook

Our work makes several predictions for the structure and temperature
profile of the protosolar disk. Assuming that the highlighted processes
are not unusual, magnetic fields, disk winds and reversed surface density profiles should commonly be found in the inner regions of protoplanetary 
disks. Advanced \hbox{3-dimensional} models can treat turbulence and viscosity
in a self-consistent manner\cite{Flock_etal_2017ApJ...835..230F},
but they do not have the ability to study effects
on small spatial scales (as needed for migration of low-mass planets)
and long time scales (as needed to understand the disk evolution).
Adaptive-optics imaging instruments (e.g., Extremely Large Telescope)
and long-baseline interferometric observatories (e.g., ALMA)
will help to resolve the inner edges of protoplanetary disks
and may eventually determine disk profiles on a sub-au scale.
These efforts will be crucial for understanding the formation of worlds
similar to our own, as well as their habitability.

\begin{methods}

\paragraph{Radiation-hydrodynamic model.}
Our system of 2D radiation hydrodynamic equations includes
the continuity of gas,
Navier--Stokes,
gas energy,
equation of state,
continuity of pebbles,
momentum of pebbles,
accretion onto protoplanets, and
equation of motion for protoplanets,
with a detailed formulation given in the SI
(or in ref.~\cite{Chrenko_etal_2017A&A...606A.114C}).
Our code is a substantial modification of {\it Fargo}\cite{Masset_2000A&AS..141..165M},
with 20 source terms, including
the viscous heating,
stellar irradiation,
vertical cooling,
accretion heating,
flux-limited diffusion approximation of the radiation transfer,
solved by the successive overrelaxation method (SOR),
two-fluid approximation,
with a pressure-less fluid for pebbles,
pebble accretion in both the Bondi and Hill regimes,
dynamic coupling between the pebble and gas disks,
aerodynamic Epstein drag on pebbles and the corresponding back-reaction on gas,
mutual gravity of the central body, protoplanets and gas disk,
and vertical damping due to density waves\cite{Tanaka_Ward_2004ApJ...602..388T}.

Among the code improvements, we incorporated
the Zhu opacity law\cite{Zhu_etal_2012ApJ...746..110Z}
to better describe the radiative disk structure in the terrestrial region.
Using Semenov opacities for dust\cite{Semenov_etal_2003A&A...410..611S}
would result in a comparable structure, albeit a bit more complex,
with even more evaporation lines.
We corrected a mistake in our torque calculations
(a minor shift of the sound speed field),
which was pronounced in the terrestrial zone,
and implemented an additional stabilisation of the SOR,
which was needed close to the hot inner edge.
Contrary to the disk in the giant-planet zone,
where the pebble size is limited by radial drift,
here it is limited by mutual collisions.
We included a simple model for pebble evaporation.
For completeness, we also included aerodynamic drag,
which would be relevant mostly for asteroid-sized bodies.
Optionally, a viscosity profile $\nu(r)$ or a temperature-dependent $\alpha$ viscosity can be used.
We perform two relaxation procedures prior to the run:
one with an outflow boundary condition
and another with a damping.
The time span of the simulations is computationally limited to ${<}\,10^5\,{\rm yr}$,
even though the algorithm is parallelized (MPI and OpenMP)
and runs on ${\simeq}\,100$ CPU cores.

\paragraph{N-body model.}
Our N-body model used to explore accretion on 
$\sim10^7\,{\rm yr}$ timescales
is based on {\it SyMBA}\cite{Duncan_etal_1998AJ....116.2067D}.
The symplectic algorithm preserves the total energy for a purely gravitational N-body system,
handles close encounters between bodies by adaptively subdividing the time step,
and efficiently detects and resolves collisions.
A number of additional processes have been included from our hydrodynamic simulations.
Hereinafter, we focus on the effect of migration,
parameterized by its time scale $\tau(m)$,
0-torque radius $r_0(m)$, and
migration rate $\dot a(a-r_0,\tau)$.
These parameters are functions of protoplanet mass~$m$.
We include torque reductions for an increased eccentricity~$e$, or
the Lindblad torque reversal which is especially important
in disks with low aspect ratios ($h = H/r \simeq 0.03$ in our case).
Eccentricity and inclination damping,
with the time scales $\tau_{\rm e}$ and $\tau_{\rm i}$,
is supplemented by hot-trail forcing,
which sets the minimum values,
denoted as $e_{\rm hot}$ and $i_{\rm hot}$.
In most simulations, we keep all these parameters constant,
neglecting a potentially complex disk evolution.
We consider them to either represent time-averaged values,
or correspond to a disk (or a part of it)
which was in steady state for a prolonged time.
The effects were implemented as additional transversal, radial or vertical accelerations,
and were inserted in the 'kick' term of the integrator (SI).

\paragraph{Initial conditions.}
There is a significant freedom in the selection of initial conditions,
especially in terms of the surface density~$\Sigma$.
The assumption of minimum-mass solar nebula
does {\em not\/} necessarily hold in the terrestrial zone
if the terrestrial planets accreted a substantial part of their mass as
pebbles or chondrules\cite{Johansen_etal_2015SciA....1E0109J},
originating at larger heliocentric distances and drifting inwards.
Drifting pebbles can represent a larger source than solids formed {\em in situ\/},
as the total amount of solid material in the solar nebula
is of the order $130\,M_{\rm E}$\cite{Levison_etal_2015Natur.524..322L}.
A filtering factor of individual terrestrial protoplanets,
in other words, a fraction of inward-drifting solid material
which is accreted by the planet, reaches a few per cent,
depending on~$\Sigma$ and~$m$.

At a later stage the disk must have been dissipating,
with substantially lower~$\Sigma$, and this may potentially produce
a very interesting dynamics of the embedded protoplanets.
For these reasons, we deliberately used disks with low~$\Sigma$ values,
which are formally much less massive than the MMSN.
On contrary, a more massive disk (as in~\cite{Levison_etal_2015Natur.524..322L})
would produce too much viscous heating and evaporation lines further out,
unless the gas disk was actually less viscous.

\end{methods}

\begin{addendum}
\item[Data availability]
The initial conditions of all simulations as well as
selected snapshots of hydrodynamical simulations
and data used to produce the respective figures
are available at
\url{http://sirrah.troja.mff.cuni.cz/~mira/fargo_terrestrial/}.

\item[Code availability]
The code {\it Thorin\/} is publicly available at
\url{http://sirrah.troja.mff.cuni.cz/~chrenko/}
(and its specific version used in this study at the previous URL).
The code {\it SyMBA\/} used in simulations is proprietary,
but its specific part implementing additional accelerations
is available.
\end{addendum}

%\bibliography{references}

\begin{addendum}
\item
The work of M.B. and O.C. has been supported by the Grant Agency of the Czech
Republic (grant no.\ 18-06083S).
The work of O.C. has been supported by Charles University
(research program no.\ UNCE/SCI/023;
project GA~UK no.\ 624119;
project SVV-260441).
D.N.'s work was supported by the NASA SSERVI and XRP programs.
Access to computing and storage facilities owned by parties and projects
contributing to the National Grid Infrastructure MetaCentrum,
provided under the programme ``Projects of Large Research, Development,
and Innovations Infrastructures'' (CESNET LM2015042),
is greatly appreciated.
The work was also supported by The Ministry of Education, Youth and Sports
from the Large Infrastructures for Research, Experimental Development
and Innovations project ``IT4Innovations National Supercomputing Center'' (LM2015070).
We are grateful to W.~F.~Bottke and A.~Morbidelli for valuable discussions.
We also thank R.~Fischer for sharing her geochemical computations with us.

\item[Competing Interests]
The authors declare that they have no competing financial interests.

\item[Correspondence]
Correspondence and requests for materials should be addressed to
M.B.\hfil\break (email: mira@sirrah.troja.mff.cuni.cz).

\end{addendum}

%%%%%%%%%%%%%%%%%%%%%%%%%%%%%%%%%%%%%%%%%%%%%%%%%%%%%%%%%%%%%%%%%%%%%%%%

\vfill\eject

\begin{figure2}
\centering
\includegraphics[width=14cm]{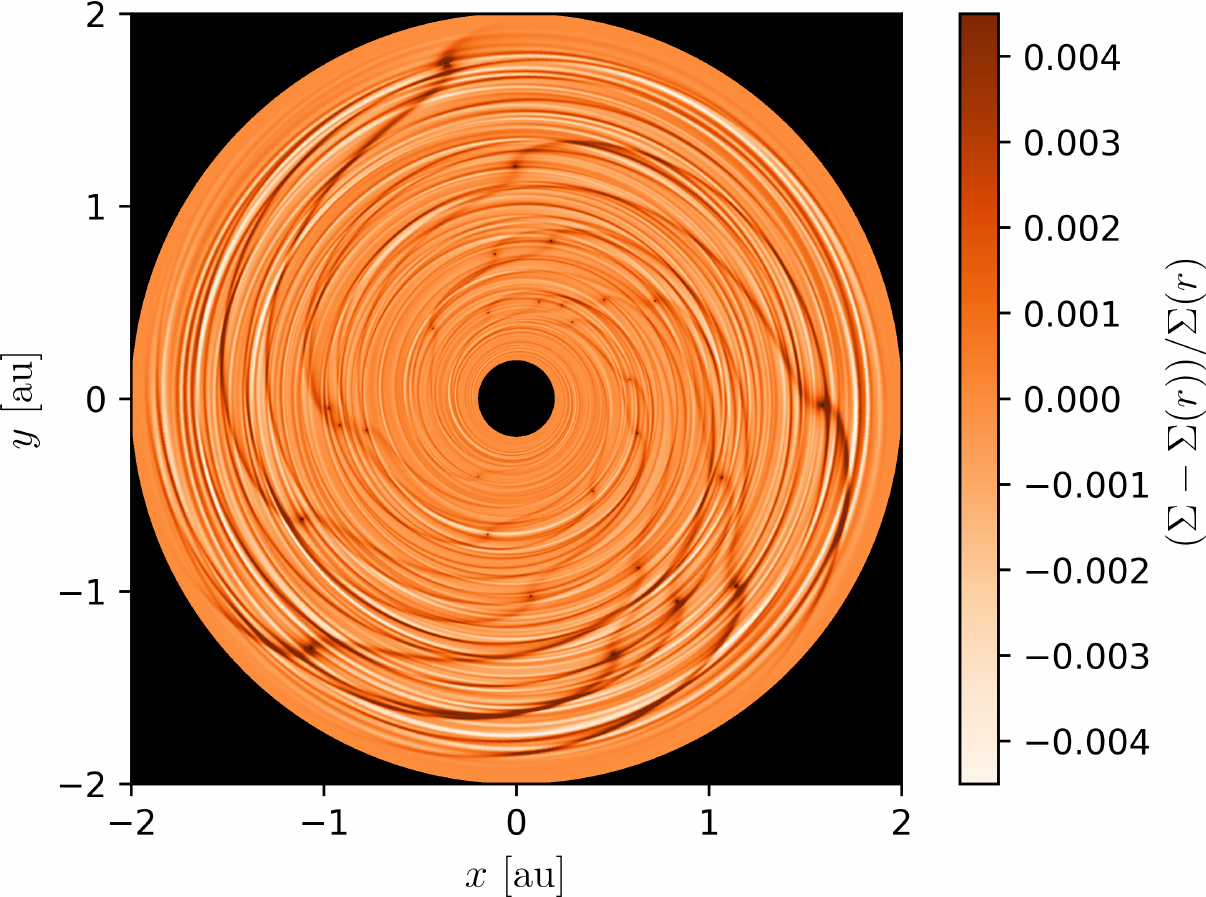}
\caption{
Radiation-hydrodynamic model of the terrestrial planet zone
with Mercury- to Mars-size protoplanets migrating in a gas disk.
To emphasize gas density perturbations induced by the protoplanets,
here we plot $(\Sigma-\Sigma(r))/\Sigma(r)$, where
$\Sigma$ is the local surface density and
$\Sigma(r)$ is the azimuthally averaged surface density at radius $r$.
This effectively removes the global radial dependence of $\Sigma$ and
highlights structures at the Lindblad resonances and corotation regions, 
which drive planetary migration.
The initial disk profile was assumed to be
$\Sigma(r) = \Sigma_0\,r^{-1.5}$,
with $\Sigma_0 = 750\,{\rm g}\,{\rm cm}^{-2}$ at $r=1$ au, i.e.
close to the Minimum Mass Solar Nebula (MMSN).
The total mass of the protoplanets was set to $2\,M_{\rm E}$.}
\label{marssize_2048}
\end{figure2}

\vfill\eject

\begin{figure2}
\centering
\includegraphics[width=14cm]{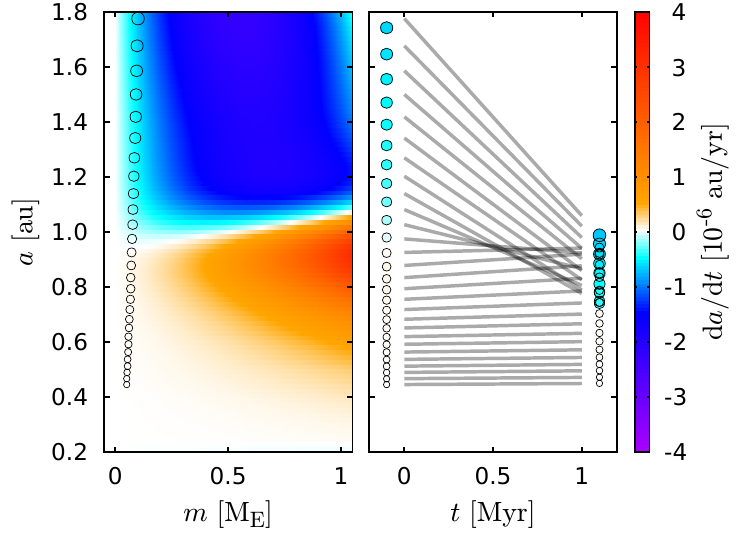}
\caption{
Convergent migration toward ${\sim}\,1\,{\rm au}$
is illustrated for Mercury- to Earth-mass protoplanets.
Here we adopted an MRI-active disk with the viscosity~$\nu(r)$
prescribed as a function of radial distance $r$.
The migration rates, ${\rm d}a/{\rm d}t$,
were computed from hydrodynamical simulations (see Figure~1),
which include the Lindblad, corotation and heating torques,
and from semianalytical formulae\cite{Paardekooper_etal_2011MNRAS.410..293P}.
Panel~(a) shows  ${\rm d}a/{\rm d}t$ as a function of
protoplanet's semimajor axis~$a$ and mass~$m$.
The convergence radius in~(a) slightly increases with mass.
The characteristic values of $|{\rm d}a/{\rm d}t|$ are
$\sim\!10^{-7}$--$10^{-6}\,{\rm au}\,{\rm yr}^{-1}$,
which implies long-range orbital drift in the disk lifetime.
Panel~(b) shows extrapolated evolutions~$a(t)$ for Mercury-
to Mars-size protoplanets from our hydrodynamical simulations.
}
\label{dadt}
\end{figure2}

\vfill\eject

\begin{figure2}
\centering
\begin{tabular}{l}
\includegraphics[width=10cm]{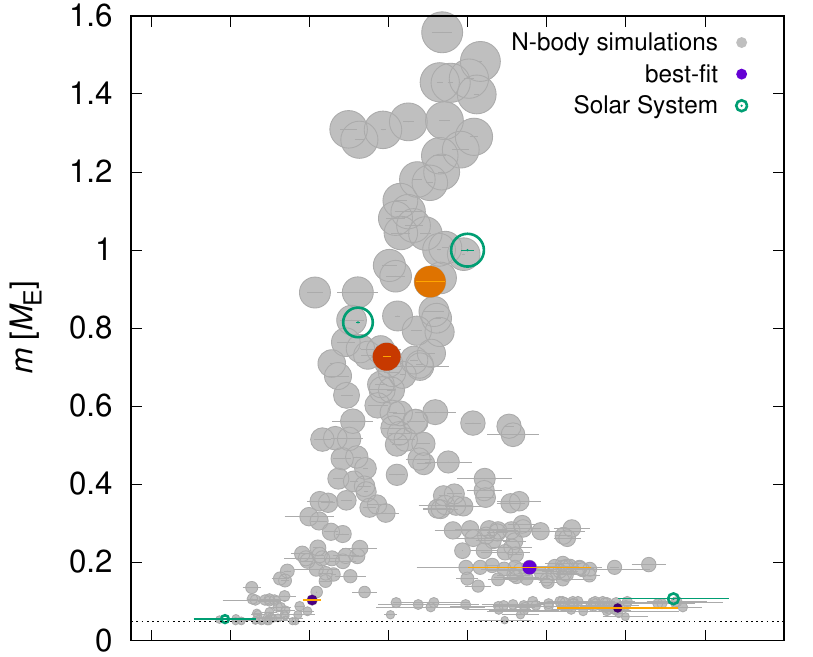} \\[-0.3cm]
\includegraphics[width=10cm]{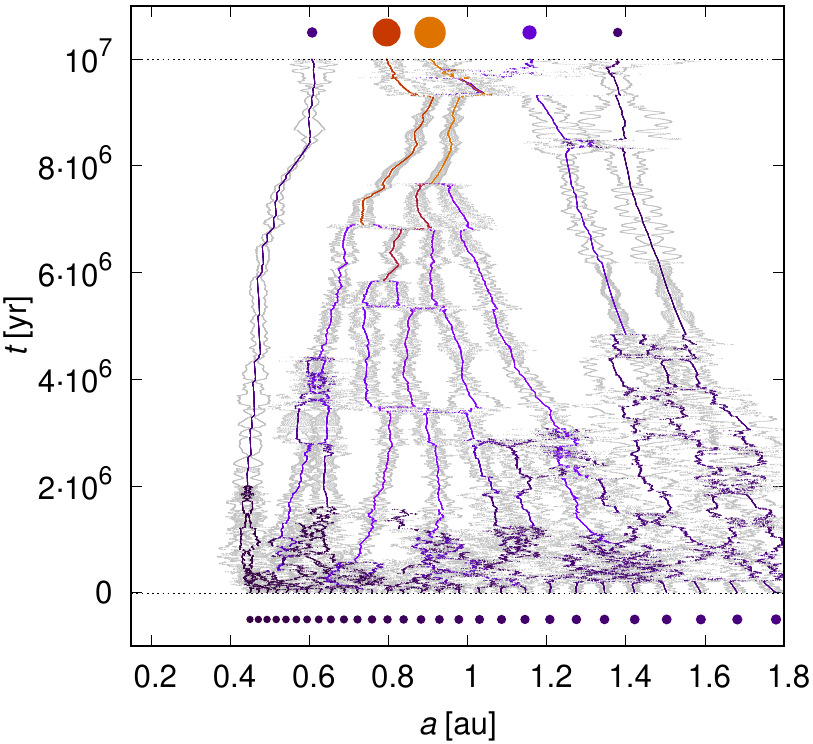}
\end{tabular}
\caption{
Convergent migration leads to a compact configuration of orbits
that matches the orbital architecture of the terrestrial planet system.
Panel~(a) shows a statistical ensemble of results from 50~individual
simulations (gray circles).
Each simulation starts with Mercury- to Mars-size protoplanets.
The protoplanetary disk lifetime is assumed to be $10^7\,{\rm yr}$.
Three to six terrestrial planets typically form in these simulations
with more massive planets near the convergence radius
and less massive planets near borders of the convergence zone 
(see SI for an in-depth analysis of the results).
The terrestrial planets are shown for reference (green circles).
Panel~(b) shows one successful simulation. The symbol sizes are
proportional to planet mass. The horizontal line segments in~(a) and 
grey lines in~(b) express the radial excursions of planets on their orbits
(i.e., measure the orbital eccentricity, $e$).
}
\label{symba11b}
\end{figure2}

\vfill\eject

\begin{figure2}
\centering
\begin{tabular}{l}
\includegraphics[width=12cm]{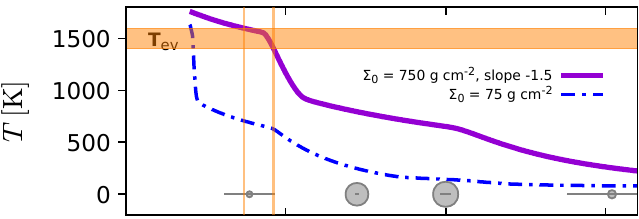} \\[-0.2cm]
\includegraphics[width=12cm]{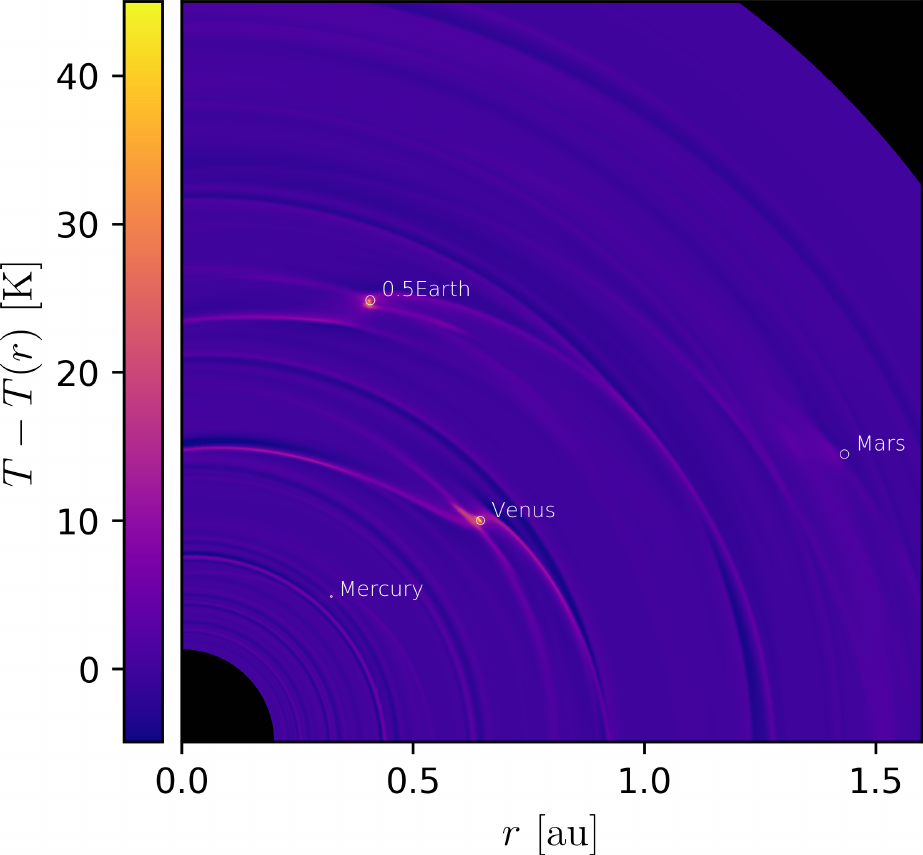}
\end{tabular}
\caption{
Temperature profile of the protoplanetary disk
determines the local chemical composition of solids,
whereas temperature perturbations affect the orbital evolution of protoplanets.
Panel~(a) shows the temperature profile~$T(r)$ for our nominal disk with
$\Sigma_0 = 750\,{\rm g}\,{\rm cm}^{-2}$ (solid violet line),
and for a dissipating disk with
$\Sigma_0 = 75\,{\rm g}\,{\rm cm}^{-2}$ (dashed blue line).
The evaporation temperature of metallic iron and Mg-rich silicates,
$T_{\rm ev}\simeq1500\,{\rm K}$, is indicated by the horizontal orange strip.
The present-day terrestrial planets are indicated by gray circles at the bottom of panel~(a).
Panel~(b) shows the temperature perturbations $\delta T=T-T(r)$ in the dissipating disk,
which arise due to accretion heating ($\delta T > 50\,{\rm K}$ for a Venus-size body).
The hot-trail effect (see hot regions behind protoplanets following their epicyclic motion)
can increase the orbital eccentricity up to $e \simeq 0.02$.
Here we highlight a case where, in addition to Mercury, Venus 
and Mars, two planets with $M=0.5\,M_{\rm E}$ were placed near 1\,au. 
This represents one of the possible configurations that may have
triggered the Moon-forming impact\cite{Canup_etal_2019}.}
\label{profiles_planets}
\label{5planet_1over10}
\end{figure2}

%%%%%%%%%%%%%%%%%%%%%%%%%%%%%%%%%%%%%%%%%%%%%%%%%%%%%%%%%%%%%%%%%%%%%%%%

\vfill\eject
\box1

\box2

\box3

\box4

%%%%%%%%%%%%%%%%%%%%%%%%%%%%%%%%%%%%%%%%%%%%%%%%%%%%%%%%%%%%%%%%%%%%%%%%

\vfill\eject

%\documentclass{nature}

%\bibliographystyle{naturemag}

%\usepackage{graphicx}
%\usepackage{amssymb}
%\usepackage{hyperref}

%\title{Terrestrial planet formation by torque-driven\\[8pt] convergent migration of planetary embryos (SI)}

%\author{M.~Bro\v z$^1$, O.~Chrenko$^1$, D.~Nesvorn\'y$^2$, N.~Dauphas$^3$ }

%\newcommand{\icarus}{Icarus}
%\newcommand{\aap}{Astron.\ Astrophys.}
%\newcommand{\aaps}{Astron.\ Astrophys.\ Suppl.\ Ser.}
%\newcommand{\nat}{Nature}
%\newcommand{\apj}{Astrophys.\ J.}
%\newcommand{\apjl}{Astrophys.\ J.\ Lett.}
%\newcommand{\mnras}{Mon.\ Not.\ R.\ Astron.\ Soc.}
%\newcommand{\grl}{Geophys.\ Res.\ Let.}
%\newcommand{\ssr}{Space Sci.\ Rev.}
%\newcommand{\aapr}{Astron.\ Astrophys.\ Rev.}

\def\d{{\rm d}}
\def\pd{\partial}
\def\vec#1{{\bf #1}}

%\begin{document}

\let\includegraphics\oldincludegraphics
\setcounter{figure}{0}%

{\Large\bfseries\noindent\sloppy\textsf{Terrestrial planet formation by torque-driven\\[-12pt] convergent migration of planetary embryos (SI)} \par}%

{\noindent M.~Bro\v z$^1$, O.~Chrenko$^1$, D.~Nesvorn\'y$^2$, N.~Dauphas$^3$}

\begin{affiliations}
 \item Institute of Astronomy, Charles University, V Hole\v sovi\v ck\'ach 2, 18000 Prague 8, Czech Republic
 \item Department of Space Studies, Southwest Research Institute, 1050 Walnut St., Suite 300, Boulder, CO 80302, USA
 \item Origins Laboratory, Department of the Geophysical Sciences and Enrico Fermi Institute, The University of Chicago, 5734 South Ellis Av., Chicago, IL 60637, USA
\end{affiliations}

\appendix

%%%%%%%%%%%%%%%%%%%%%%%%%%%%%%%%%%%%%%%%%%%%%%%%%%%%%%%%%%%%%%%%%%%%%%%%

\section{Radiation hydrodynamic model}

Our current model includes the following set of integro-differential equations
(see also \cite{Chrenko_etal_2017A&A...606A.114C}):
\begin{equation}
{\pd\Sigma\over\pd t} + \vec v\cdot\nabla\Sigma = -\Sigma\nabla\cdot\vec v - \left({\pd\Sigma\over\pd t}\right)_{\!\rm acc},\label{eq:dSigma_dt}
\end{equation}
\begin{equation}
{\pd\vec v\over\pd t} + \vec v\cdot\nabla\vec v = -{1\over\Sigma}\nabla P + {1\over\Sigma}\nabla\!\cdot\!{\sf T} - {\int\rho\nabla\!\phi\,\d z\over\Sigma} + {\Sigma_{\rm p}\over\Sigma}{\Omega_{\rm K}\over\tau}(\vec u-\vec v)\,,
\end{equation}
\begin{eqnarray}
{\pd E\over\pd t} + \vec v\cdot\nabla E &=& -E\nabla\cdot\vec v - P\nabla\cdot\vec v + Q_{\rm visc} + {2\sigma T_{\rm irr}^4\over\tau_{\rm eff}} - {2\sigma T^4\over\tau_{\rm eff}}+ \nonumber\\
&& +\, {2H\nabla\cdot{{16\sigma\lambda_{\rm lim}\over\rho_0\kappa_{\rm R}}}T^3\nabla T} + \sum_i{GM_i\dot M_i\over R_i S_{\rm cell}}\delta(\vec r-\vec r_i)\,,
\end{eqnarray}
\begin{equation}
P = \Sigma {RT\over\mu} = (\gamma-1)E\,,
\end{equation}
\begin{equation}
{\pd\Sigma_{\rm p}\over\pd t} + \vec u\cdot\nabla\Sigma_{\rm p} = -\Sigma_{\rm p}\nabla\cdot\vec u - \left({\pd\Sigma_{\rm p}\over\pd t}\right)_{\!\rm acc} - \left({\pd\Sigma_{\rm p}\over\pd t}\right)_{\!\rm evap},
\end{equation}
\begin{equation}
{\pd\vec u\over\pd t} + \vec u\cdot\nabla\vec u = -{\int\rho_{\rm p}\nabla\!\phi\,\d z\over\Sigma_{\rm p}} - {\Omega_{\rm K}\over\tau}(\vec u-\vec v)\,,
\end{equation}
\begin{equation}
\dot M_i = \int\!\!\!\int \left[\left({\pd\Sigma\over\pd t}\right)_{\!\rm acc}\! + \left({\pd\Sigma_{\rm p}\over\pd t}\right)_{\!\rm acc}\,\right] r\d\theta\d r\quad\hbox{for }\forall i\,,
\end{equation}
\begin{eqnarray}
\ddot{\vec r_i} &=& -{GM_\star\over r_i^3}\vec r_i
- \sum_{j\ne i}{GM_j\over|\vec r_i-\vec r_j|^3}(\vec r_i-\vec r_j)
+ \int\!\!\!\int\!\!\!\int{\rho\nabla\!\phi_i\,\d z\over M_i}r\d\theta\d r \,+ \nonumber\\
&&
+\, f_z\kern1pt\hat z
- {1\over2}C{\pi R_i^2\over M_i}\rho |\dot{\vec r}_i-\vec v_{\rm cell}| (\dot{\vec r}_i-\vec v_{\rm cell})
+ \int\!\!\!\int \left[\vec v\left({\pd\Sigma\over\pd t}\right)_{\!\rm acc}\!\!\! + \vec u\left({\pd\Sigma_{\rm p}\over\pd t}\right)_{\!\rm acc}\,\right] r\d\theta\d r\nonumber\\
&& \hbox{for }\forall i\,,\label{eq:ddot_r}
\end{eqnarray}
where
$\Sigma$~is the gas surface density,
$\vec v$~gas velocity,
$(\pd\Sigma/\pd t)_{\rm acc}$~gas accretion term\cite{Kley_1999MNRAS.303..696K},
$P$~vertically integrated pressure,
$\sf T$~viscous stress tensor,
$\rho = \Sigma/(\sqrt{2\pi}H)\exp[-z^2/(2H^2)]$~gas volumetric density,
$H$~scale height,
$\phi = \phi_\star + \sum\phi_i$~gravitational potential of the Sun and protoplanets,
with a cubic smoothing due to a finite cell size\cite{Klahr_Kley_2006A&A...445..747K} (not~$H$),
$z$~vertical coordinate,
$\Sigma_{\rm p}$~pebble surface density,
$\vec u$~pebble velocity,
$(\pd\Sigma_{\rm p}/\pd t)_{\rm acc}$~pebble accretion term for the Bondi and Hill regimes\cite{Lambrechts_Johansen_2012A&A...544A..32L},
$\Omega_{\rm K}$~the Keplerian angular velocity,
$\tau$~the Stokes number of pebbles,
$E$~gas internal energy,
$Q_{\rm visc}$~viscous heating term\cite{Mihalas_1984frh..book.....M},
$\sigma$~the Stefan--Boltzmann constant,
$T_{\rm irr}$~irradiation temperature\cite{Chiang_Goldreich_1997ApJ...490..368C},
$\tau_{\rm eff}$~effective optical depth\cite{Hubeny_1990ApJ...351..632H},
$T$~gas temperature,
$\lambda_{\rm lim}$~flux limiter\cite{Kley_1989A&A...208...98K},
$\rho_0$~midplane density,
$\kappa_{\rm R}$~the Rosseland opacity,
$G$~gravitational constant,
$M_i$~protoplanet mass,
$R_i$~protoplanet radius,
$S_{\rm cell}$~cell area in which it is located,
$R$~gas constant,
$\mu$~mean molecular weight,
$\gamma$~adiabatic index,
$\ddot{\vec r}_i$~acceleration of the body~$i$,
with the smoothing applied for the 3rd term;
$f_z$~vertical damping\cite{Tanaka_Ward_2004ApJ...602..388T}
(while the horizontal damping is computed self-consistently),
$C$~drag coefficient, and
$\vec v_{\rm cell}$~cell velocity.

As the numerical code is based on Fargo\cite{Masset_2000A&AS..141..165M},
it also uses the Fargo algorithm to overcome the time step
limitation due to the Keplerian motion at the inner edge of the disk.
For the numerical integration of planetary orbits we use the IAS15 integrator
from the Rebound package\cite{Rein_Spiegel_2015MNRAS.446.1424R},
with an adaptive time step to precisely handle close encounters.

The pebble evaporation term is assumed simply as:
\begin{equation}
\left({\pd \Sigma_{\rm p}\over\pd t}\right)_{\rm evap} = f_{\rm ev} \Sigma_{\rm p} {\cal H}(T-T_{\rm ev})\,,
\end{equation}
where 
$T_{\rm ev}$~denotes the evaporation temperature,
$f_{\rm ev}$~the respective rate, and
${\cal H}$~the Heaviside step function.
We use this approximation only for a single evaporation line
and pebble surface area, instead of chemical equilibrium
and Hertz--Knudsen equations.
The remaining terms were described in more detail in
\cite{Chrenko_etal_2017A&A...606A.114C} or \cite{Broz_etal_2018A&A...620A.157B}.

%%%%%%%%%%%%%%%%%%%%%%%%%%%%%%%%%%%%%%%%%%%%%%%%%%%%%%%%%%%%%%%%%%%%%%%%

\section{Limitations of the model}

Every model has its own limitations.
The Planck opacity $\kappa_{\rm P}$, which is hidden in $\tau_{\rm eff}$,
is assumed to be the same as the Rosseland opacity~$\kappa_{\rm R}$ for simplicity.
These opacities may differ substantially for temperatures
$T \simeq 1000\,{\rm K}$\cite{Malygin_etal_2014A&A...568A..91M},
in particular in stratified 3D disks with hot atmospheres,
but it is not our case, except possibly at the inner edge of the disk.

We assume the drag is always in the Epstein regime;
the Stokes regime would be necessary for pebble sizes larger than the mean-free path.
The condition $2/9 D_{\rm p} > \ell$ can be fulfilled
only in the innermost part of the most massive disk,
where $\ell = \mu m_{\rm H}/(\sigma\rho) \simeq 0.4\,{\rm cm}$.
Otherwise, it should not affect our simulations.
We also use no reduction factor of the accretion rate\cite{Ormel_Klahr_2010A&A...520A..43O,Ida_etal_2016A&A...591A..72I},
which would be needed for $\tau \gg 1$.

\begin{figure}
\centering
\begin{tabular}{cc}
\includegraphics[width=9cm]{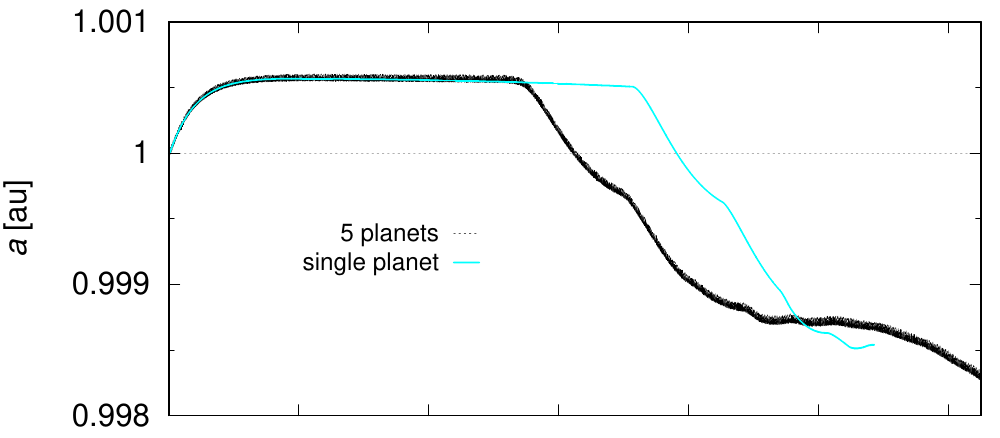} \\
\includegraphics[width=9cm]{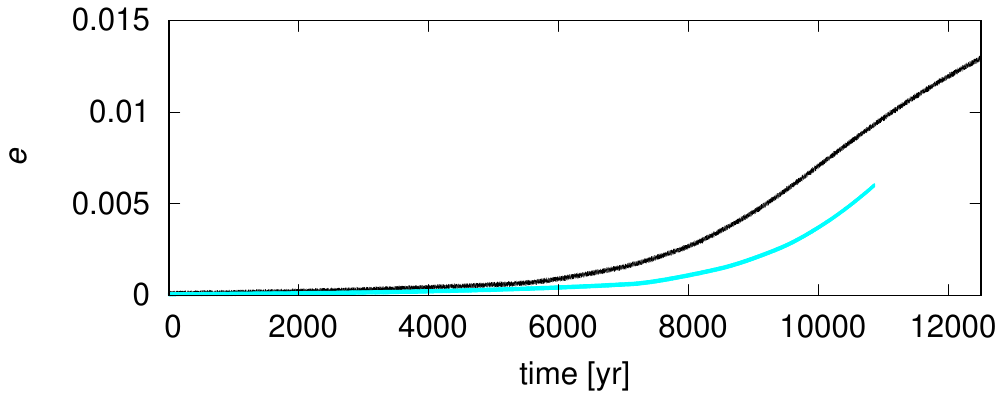} \\
\end{tabular}
\caption{Evolution of semimajor axis~$a$ (bottom) and eccentricity~$e$ (top)
computed for an {\em isolated\/} 0.5~Earth-mass body (cyan),
in comparison with the same body affected by another 4~protoplanets
in the terrestrial region (black). The migration rate is initially
the same in both cases, with only minor oscillations in the latter,
due to secular perturbations by distant planets.
The long-term evolution of eccentricity due to the hot-trail effect is similar,
except the single planet exhibits a slower increase of~$e$.
The reason is the coupling of secular perturbations (with terms proportional to~$e$)
with the hot-trail, which in turn produces a faster increase of~$e$.
The evolution of $a(t)$ is then also affected by~$e$.
}
\label{test_earth_nbody.orbits.at}
\end{figure}

In a complex model ($9+4N$ scalar integro-differential equations in our case;
$N$ denotes the number of protoplanets)
it may be sometimes difficult to recognise which terms induce the behaviour
of interest, because all terms are interrelated (by means of Eqs.~(\ref{eq:dSigma_dt}) to (\ref{eq:ddot_r})).
For example, a question might be, whether the increase of planetary eccentricities~$e$
is driven by hydrodynamic phenomena or secular perturbations (or both).
For this purpose we are sometimes forced to use simplified models
with selected terms switched off, or a limited number of protoplanets ($N = 1$).
An example is shown in Figure~\ref{test_earth_nbody.orbits.at},
which confirms that $e$~is increased due to the hot-trail effect.

\begin{figure}
\centering
\includegraphics[width=9cm]{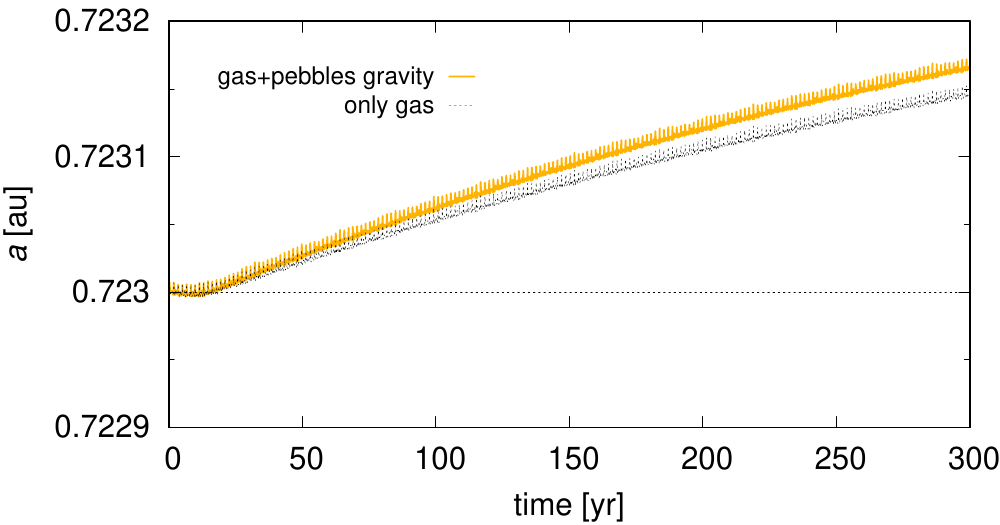}
\caption{Semimajor axis~$a$ vs time~$t$ computed for a Venus-mass body
affected by the gravity of the gas disk only (dotted)
and the gravity of gas plus pebbles (yellow).
The difference in the migration rate is at most 10\,\%
which is considered negligible in our context.
}
\label{test_pebblegrav_nbody.orbits.at_2}
\end{figure}

Although we do compute the gravity of planets acting on pebbles,
the gravity of pebbles acting on planets\cite{Benitez_Pessah_2018ApJ...855L..28B} is not included
in most of our simulations, because the pebble flux~$\dot M_{\rm p}$
and the pebble-to-gas ratio $\Sigma_{\rm p}/\Sigma \simeq 10^{-3}$ to $10^{-2}$
are generally low, as well as the Stokes number $\tau \simeq 0.01$.
For a $M = 1\,M_{\rm E}$ planet on a circular fixed orbit,
without pebble accretion and without back-reaction on gas,
the ratio of torques would be $\Gamma_{\rm p}/\Gamma_{\rm g} \simeq 0.01$ to $0.1$
according to \cite{Benitez_Pessah_2018ApJ...855L..28B},
which can be considered negligible.
Because the situation is actually more complicated --- protoplanets in our simulations
are generally eccentric, not fixed, $\Sigma_{\rm p}$ is decreased by accretion,
and the gas flow~$\vec v$ is changed by the pebble flow~$\vec u$ ---
we checked that the migration rates do {\em not\/} change substantially
when the respective term $\int\!\!\int\!\!\int{\rho_{\rm p}\!\nabla\!\phi_i\,\over M_i} \d z\,r\d\theta\d r$
is added to Eq.~(\ref{eq:ddot_r}); see Figure~\ref{test_pebblegrav_nbody.orbits.at_2}.

We assume that all pebbles have the same chemical composition,
and we do not track their composition explicitly.
For simplicity, we also use a single evaporation temperature for pebbles,
corresponding to that of high-temperature minerals (iron, orthopyroxene, olivine).
Nevertheless, the integral opacities~$\kappa_{\rm R}$ contain dust
evaporation implicitly, for several distinct minerals
(water ice, graphite, and remaining grains in case of \cite{Zhu_etal_2012ApJ...746..110Z};
and even more minerals in case of \cite{Semenov_etal_2003A&A...410..611S}).
We can thus `guess' the chemical composition of pebbles from to the local value of~$\kappa_{\rm R}$.
The only problem is that we overestimate the pebble fluxes
on protoplanets orbiting interior to the evaporation lines,
but this is somewhat suppressed by the fact that $M_{\rm p}$
is a free parameter and we can optionally use a lower value.

% term $+(\pd\Sigma/\pd t)_{\rm acc}$ in gas continuity
% recondensation of pebbles
% Jupiter, Saturn formation
% low-\Sigma phase may not occur, if the dissipation is inside-out
% constant density of protoplanets

%%%%%%%%%%%%%%%%%%%%%%%%%%%%%%%%%%%%%%%%%%%%%%%%%%%%%%%%%%%%%%%%%%%%%%%%

\section{Parameters and initial profiles}

The MRI-active disk is defined by
its gas flux $\dot M = 10^{-8}\,M_\odot\,{\rm yr}^{-1}$
and the kinematic viscosity $\nu(r) = 1.1\times 10^{14}\,{\rm cm}^2\,{\rm s}^{-1}\,(r/1\,{\rm au})^s$,
where the exponent changes smoothly,
$s = -2 + 2.5\{1-\tanh[(1\,{\rm au}-r)/(0.15\,{\rm au})]\}/2$,
in order to obtain a~surface density profile~$\Sigma(r)$
similar to a~layered-accretion disk\cite{Kretke_Lin_2012ApJ...755...74K}.
The nominal disk, similar to the MMSN\cite{Hayashi_1981PThPS..70...35H},
is defined by the surface density
$\Sigma(r) = \Sigma_0 (r/1\,{\rm au})^{-1.5}$,
with $\Sigma_0 = 750\,{\rm g}\,{\rm cm}^{-2}$,
and constant $\nu = 1.5\times 10^{14}\,{\rm cm}^2\,{\rm s}^{-1} \doteq 3.3\times10^{-5}$ [code units].
This value is compatible with our previous paper\cite{Chrenko_etal_2017A&A...606A.114C}
focussing on the giant-planet zone, but it is 3~times lower
because~$\nu$ is often parametrized as\cite{Shakura_Sunyaev_1973A&A....24..337S}
$\nu = \alpha c_{\rm s} H$,
where
$\alpha$~is yet another free parameter,
$c_{\rm s}$~the sound speed, and
$H$~the vertical scale height.
The radial profiles of the last two quantities led us to use lower~$\nu$ at 1\,au.
Alternatively, we may use $\alpha$ parameter directly,
which would slightly change the profiles.
The values corresponding to $\nu$ above would be
$\alpha \simeq 0.001$ to $0.005$.

The common parameters of the simulations are as follows:
adiabatic index $\gamma = 1.4$,
molecular weight $\mu = 2.4\,{\rm g}\,{\rm mol}^{-1}$,
disc albedo $A = 0.5$,
vertical opacity drop $c_\kappa = 0.6$,
effective temperature of the star $T_\star = 4370\,{\rm K}$,
stellar radius $R_\star = 1.5\,R_\odot$,
softening parameter is~$0.5R_{\rm H}$.
The entire Hill sphere is considered when calculating disk$\,\leftrightarrow\,$planet interactions.
The inner boundary is $r_{\rm min} = 0.2\,{\rm au}$,
outer boundary $r_{\rm max} = 2.0\,{\rm au}$,
a~damping boundary condition (e.g. \cite{Kley_Dirksen_2006A&A...447..369K}) is used to prevent spurious reflections,
and applied up to $1.2r_{\rm min}$ and from $0.9r_{\rm max}$ on;
the vertical damping parameter is~0.3,
pebble flux $\dot M_{\rm p} = 2\times10^{-4}\,M_{\rm E}\,{\rm yr}^{-1}$,
turbulent stirring parameter $\alpha_{\rm p} = 10^{-4}$,
which determines the scale height of the pebble disk, $H_{\rm p}/H = \sqrt{\alpha_{\rm p}/\tau}$,
the Schmidt number ${\rm Sc} = 1$,
pebble coagulation efficiency $\epsilon_{\rm p} = 0.5$,
pebble bulk density $\rho_{\rm p} = 3\,{\rm g}\,{\rm cm}^{-3}$,
fragmentation factor $f_{\rm f} = 1.0$,
fragmentation threshold $10\,{\rm m}\,{\rm s}^{-1}$~\cite{Birnstiel_etal_2012A&A...539A.148B},
turbulence parameter $\alpha_{\rm t} = 10^{-3}$,
evaporation temperature $T_{\rm ev} = 1500\,{\rm K}$~\cite{Pollack_etal_1994ApJ...421..615P},
evaporation rate $f_{\rm ev} = 10^{-3}$,
embryo density $\rho_{\rm em} = 3\,{\rm g}\,{\rm cm}^{-3}$ (constant).

The spatial discretisation we nominally use is $1024\times 1536$ cells,
either with arithmetic spacing, or logarithmic if we need to improve
the radial resolution for low-mass embryos.
See Appendix~\ref{sec:convergence} for a convergence test.
The discretisation in time is controlled by the CFL condition;
the maximal time step is $\Delta t = 0.314159\,[{\rm c.u.}] = 1/20\,P_{\rm orb}$ at 1.0\,au.
Orbital elements are output every $20\,\Delta t$,
and hydrodynamical fields every $500\,\Delta t$.
The nominal time span is approximately 8\,kyr.
The relative precision of the IAS15 integrator\cite{Rein_Spiegel_2015MNRAS.446.1424R}
is set to $10^{-9}$.

\begin{figure}
\centering
\includegraphics[width=8cm]{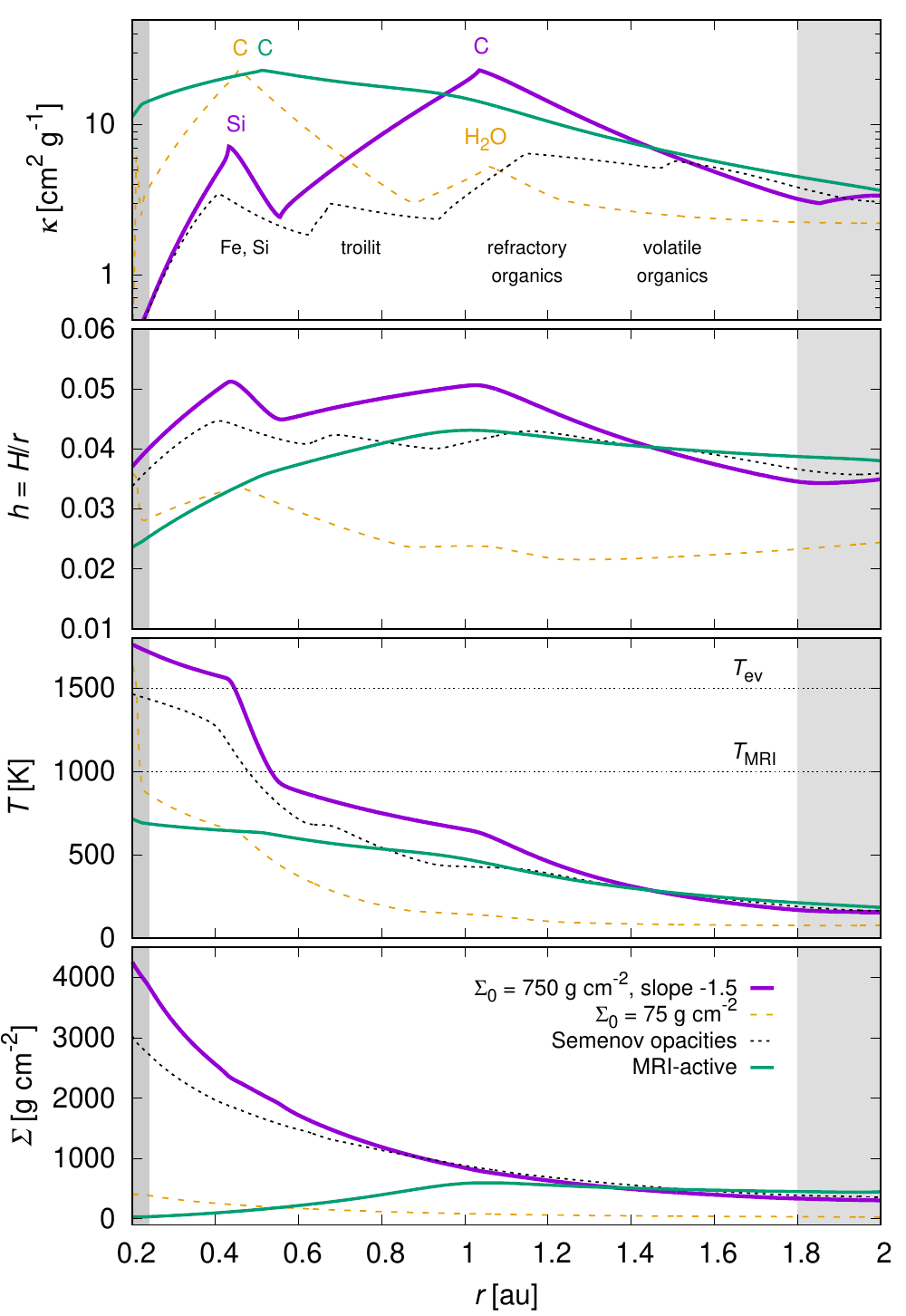}
\caption{
Initial radial profiles of the gas disk, namely
the surface density~$\Sigma$ (bottom),
temperature~$T$,
aspect ratio~$h = H/r$, and
Rosseland opacity~$\kappa$ (top).
They are shown for 4~variants:
the nominal disk with $\Sigma_0 = 750\,{\rm g}\,{\rm cm}^{-2}$,
the dissipating disk ($\Sigma_0 = 75\,{\rm g}\,{\rm cm}^{-2}$),
the nominal one with different opacities\cite{Semenov_etal_2003A&A...410..611S,Malygin_etal_2014A&A...568A..91M},
and the MRI-active disk with the prescribed kinematic viscosity~$\nu(r)$.
The temperature
$T_{\rm MRI}$ indicates a possible value for the onset of magneto-rotational instability, and
$T_{\rm ev}$ for the evaporation of (all) pebbles.
}
\label{profiles_kappa2}
\end{figure}

\begin{figure}
\centering
\includegraphics[width=8cm]{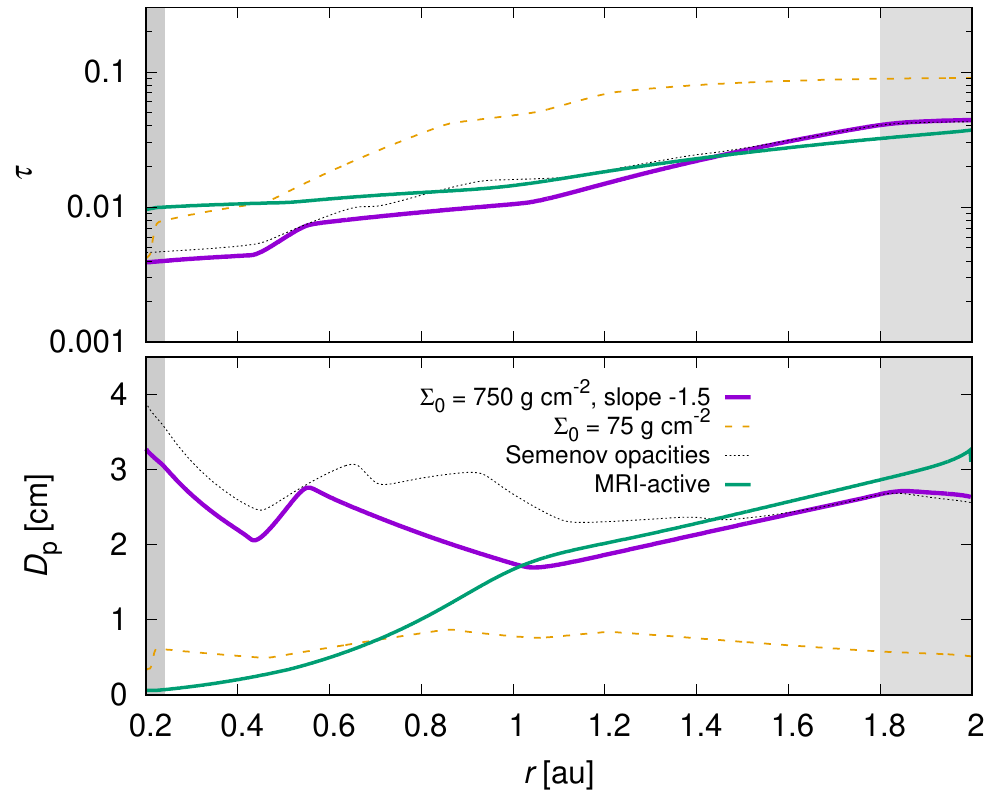} \\
\caption{
Initial radial profiles of the pebble disk.
The pebble size~$D_{\rm p}$ (bottom)
and the Stokes number~$\tau$ (top)
are shown for 4~variants:
the nominal disk with $\Sigma_0 = 750\,{\rm g}\,{\rm cm}^{-2}$,
the dissipating disk ($\Sigma_0 = 75\,{\rm g}\,{\rm cm}^{-2}$),
the nominal one with different opacities\cite{Semenov_etal_2003A&A...410..611S,Malygin_etal_2014A&A...568A..91M},
and the MRI-active disk.
}
\label{profiles_pebble2}
\end{figure}

Initial profiles of the gas disk are shown in Figure~\ref{profiles_kappa2}.
The respective profiles of the pebble disk are shown in Figure~\ref{profiles_pebble2}.
In the terrestrial region their sizes are usually limited
by fragmentation\cite{Birnstiel_etal_2012A&A...539A.148B}
and they reach a few centimetres. The Stokes numbers are
in the range $\tau \simeq 0.01$ to $0.1$, corresponding
to a partial coupling between gas and pebbles.

%%%%%%%%%%%%%%%%%%%%%%%%%%%%%%%%%%%%%%%%%%%%%%%%%%%%%%%%%%%%%%%%%%%%%%%%

\section{Supplementary figures}

\begin{figure}
\centering
\includegraphics[width=14cm]{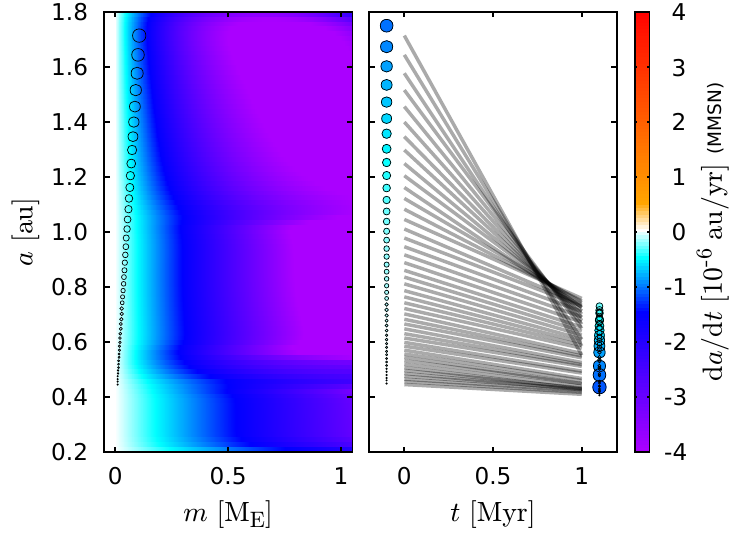}
\caption{
Migration rates ${\rm d}a/{\rm d}t$ in a gas disk
with the surface density $\Sigma = 750\,{\rm g}\,{\rm cm}^{-2}\,r^{-1.5}$,
i.e. close to the minimum-mass solar nebula (MMSN\cite{Hayashi_1981PThPS..70...35H}).
The dependence on the semimajor axis~$a$ and protoplanet mass~$m$ (\cite{Paardekooper_etal_2011MNRAS.410..293P}; left)
and an extrapolated evolution $a(t)$ for lunar- to Mars-size protoplanets
($0.01$ to $0.1\,M_{\rm E}$; right) are shown.
}
\label{dadt21_mmsn}
\end{figure}

As a supplement to Fig.~2 from the main text, we show migration rates in the MMSN disk in Figure~\ref{dadt21_mmsn}.
The migration rates of low-mass protoplanets are all negative
and the convergence zone does not formally exist.
However, these rates do depend on protoplanet mass and they are sufficiently different,
${\rm d}a/{\rm d}t \doteq -10^{-7}$ to \hbox{$-10^{-6}\,{\rm au}\,{\rm yr}^{-1}$},
such that lunar- and Mars-sized protoplanets approach each other
and also concentrate in a narrow annulus.
However, the convergence may not be so pronounced if protoplanets
do not follow a simple radial dependence of the isolation mass\cite{Walsh_Levison_2019Icar..329...88W}.

Alternatively, for wind-driven disks\cite{Ogihara_etal_2018A&A...612L...5O}
the viscosity in the midplane can be very low,
which would also break the convergence zone
because the corotation torque is not kept unsaturated.
However, we can assume the viscous stress is provided
by non-ideal MHD effects\cite{McNally_etal_2018JPhCS1031a2007M}.

\begin{figure}
\centering
\includegraphics[width=8cm]{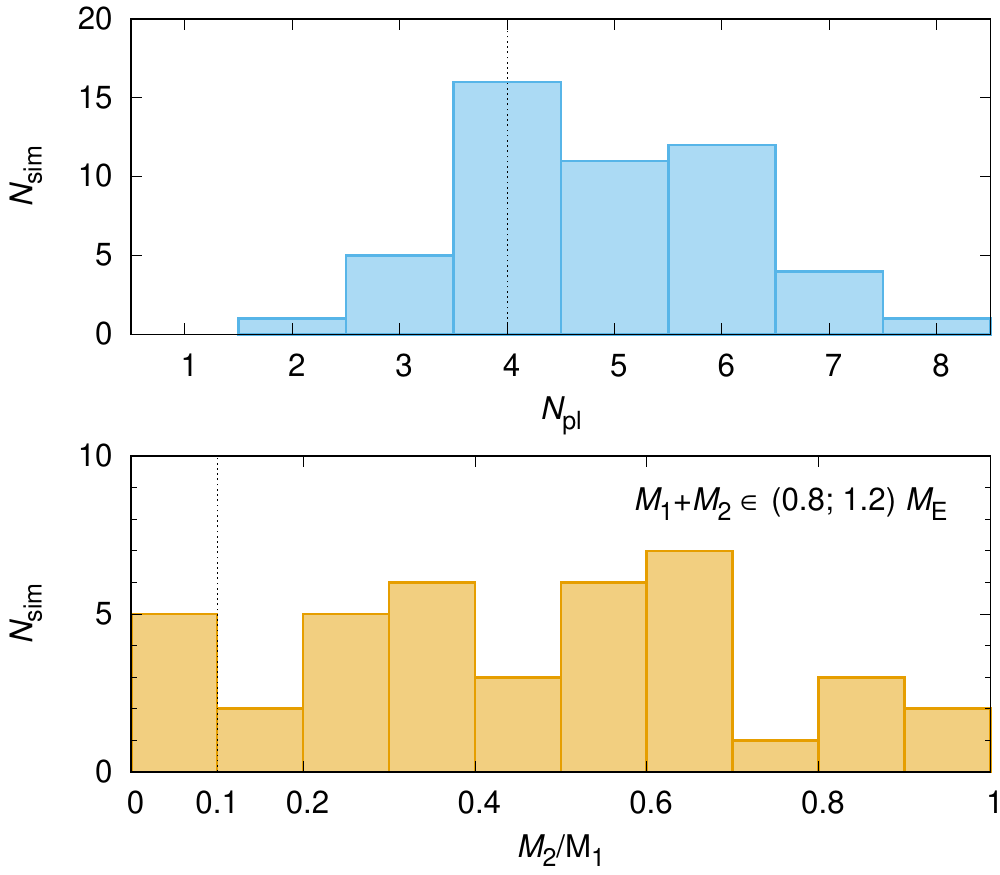}
\caption{Distribution of the number of planets $N_{\rm pl}$ (top)
and projectile-to-target mass ratios $M_2/M_1$
for Moon-forming impacts (with $M_1+M_2 \in (0.8; 1.2)\,M_{\rm E}$) (bottom),
in our set of N-body simulations with convergent migration.}
\label{hist2_symba11b_reductions5}
\end{figure}

As a supplement to Fig.~3, we show the distribution of
the number of planets in our set of N-body simulations,
as well as the projectile-to-target mass ratios of Moon-forming
impacts in Figure~\ref{hist2_symba11b_reductions5}.

\begin{figure}
\centering
\includegraphics[width=10cm]{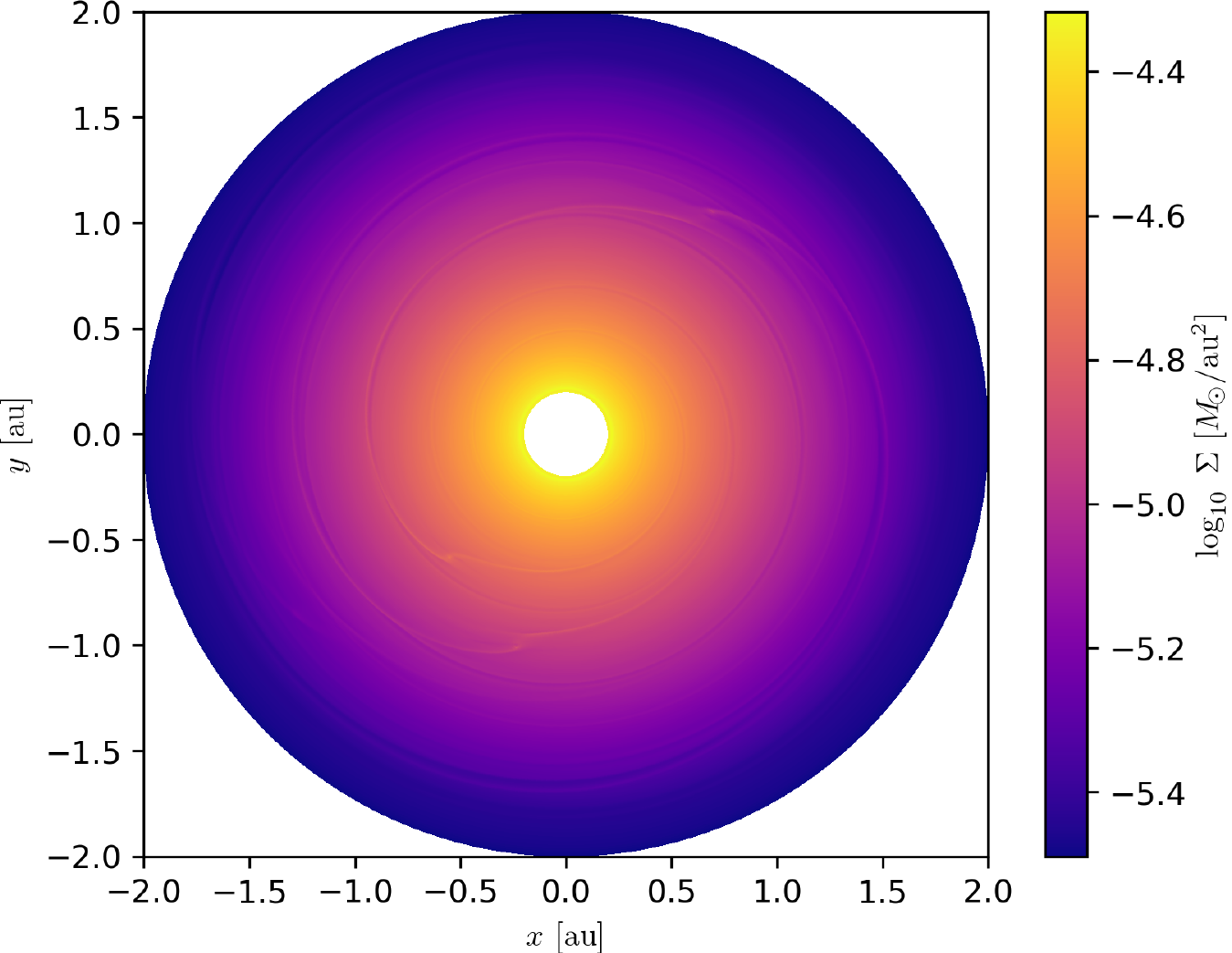}
\includegraphics[width=10cm]{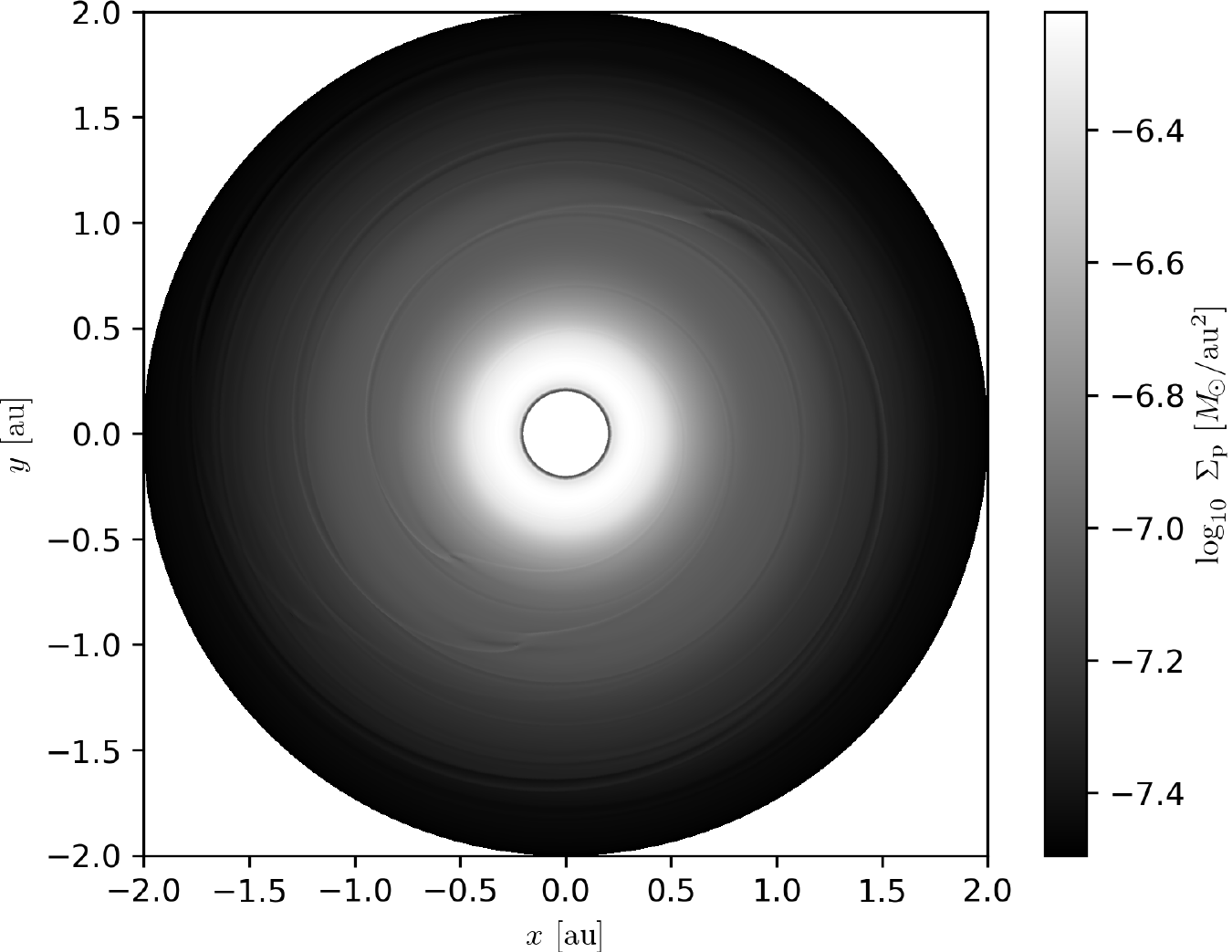}
\caption{
Surface density of gas~$\Sigma$ (left) and pebbles~$\Sigma_{\rm p}$ (right)
of the dissipating disk with the initial value of $\Sigma_0 = 75\,{\rm g}\,{\rm cm}^{-2}$.
It corresponds to Fig.~4 from the main text (for $t = 0.1\,{\rm kyr}$),
but here we show the situation at the end of simulation ($t = 24\,{\rm kyr}$).
The structures in both disks can be seen.
The hot-trail effect is well developed and the protoplanets gained substantial
non-zero eccentricities this way (up to $e \simeq 0.02$).
Another contributing factor is the mean-motion resonance 4:3 between protoplanets~4 and~5.
The pebble surface density~$\Sigma_{\rm p}$ decreases in the vicinity of each planet
due to ongoing pebble accretion.
}
\label{gasdens960_3}
\end{figure}

\begin{figure}
\centering
\includegraphics[width=10cm]{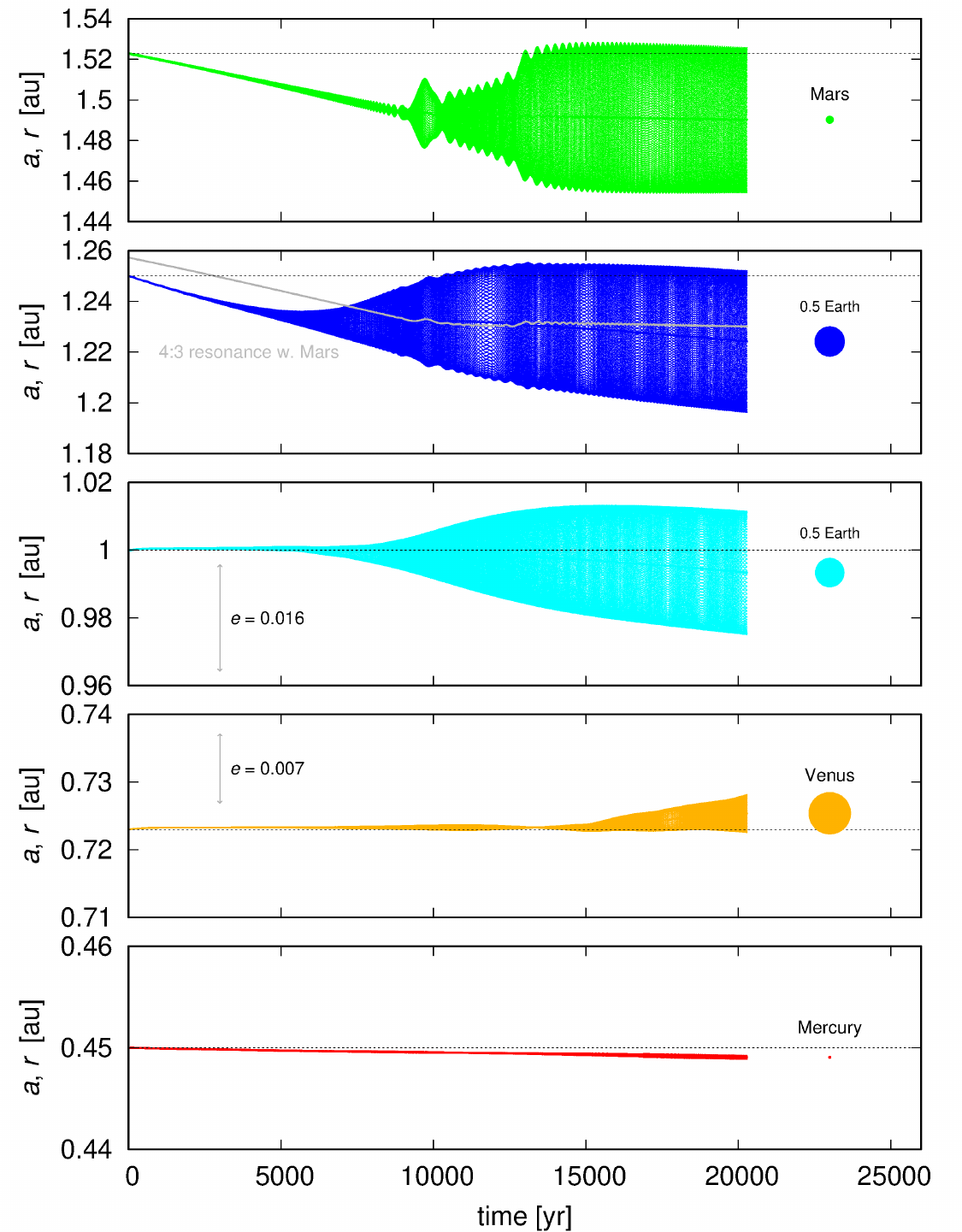}
\caption{
Short-term orbital evolution of 5~terrestrial planets
in the cold dissipating disk ($\Sigma_0 = 75\,{\rm g}\,{\rm cm}^{-2}$),
as computed by the radiation hydrodynamic (RHD) model.
The semimajor axis~$a$ and heliocentric distance~$r$ vs time~$t$ are shown;
the oscillations of $r$ correspond to the eccentricity~$e$.
The increase of eccentricities by the hot-trail effect up to $e \simeq 0.02$
is comparable to the osculating eccentricities of Venus and Earth,
$e = 0.007$ and $0.016$ respectively (gray arrows).
At $t \doteq 8000\,{\rm yr}$ a temporary capture in the 4:3 mean-motion
resonance between protoplanets 4 and 5 occurs.
Delivery of water by icy pebbles is possible in this situation
because the snowline is at approximately 1\,au.}
\label{nbody.orbits.at3}
\end{figure}

In Figure~\ref{gasdens960_3}, we show both gas and pebble surface
densities in the cold dissipating disk ($\Sigma_0 = 75\,{\rm g}\,{\rm cm}^{-2}$), at the end of the respective simulation.
Finally, we show the corresponding short-term orbital evolution
demonstrating the hot-trail effect, as computed by our
hydrodynamical model (Figure~\ref{nbody.orbits.at3}).

%%%%%%%%%%%%%%%%%%%%%%%%%%%%%%%%%%%%%%%%%%%%%%%%%%%%%%%%%%%%%%%%%%%%%%%%

\section{A convergence test for a Mercury-size body}\label{sec:convergence}

\begin{figure}
\centering
\begin{tabular}{cc}
Mercury & Mars \\
\includegraphics[width=8cm]{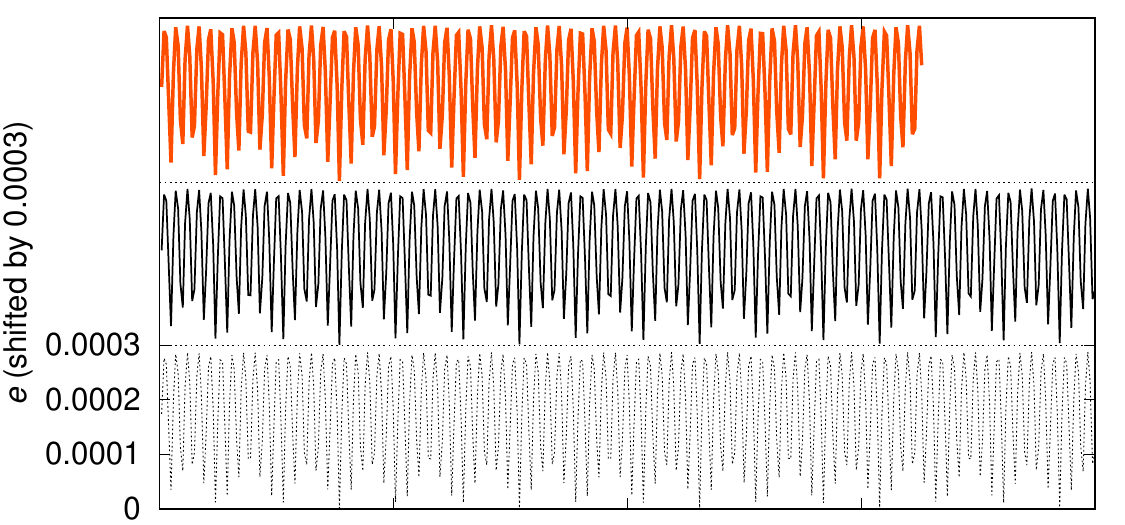} &
\includegraphics[width=8cm]{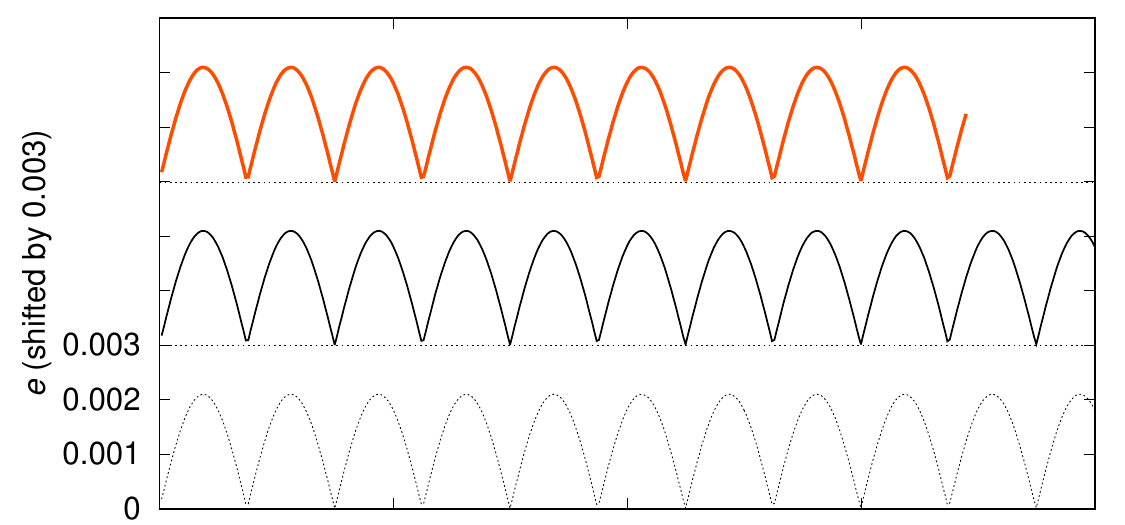} \\
\includegraphics[width=8cm]{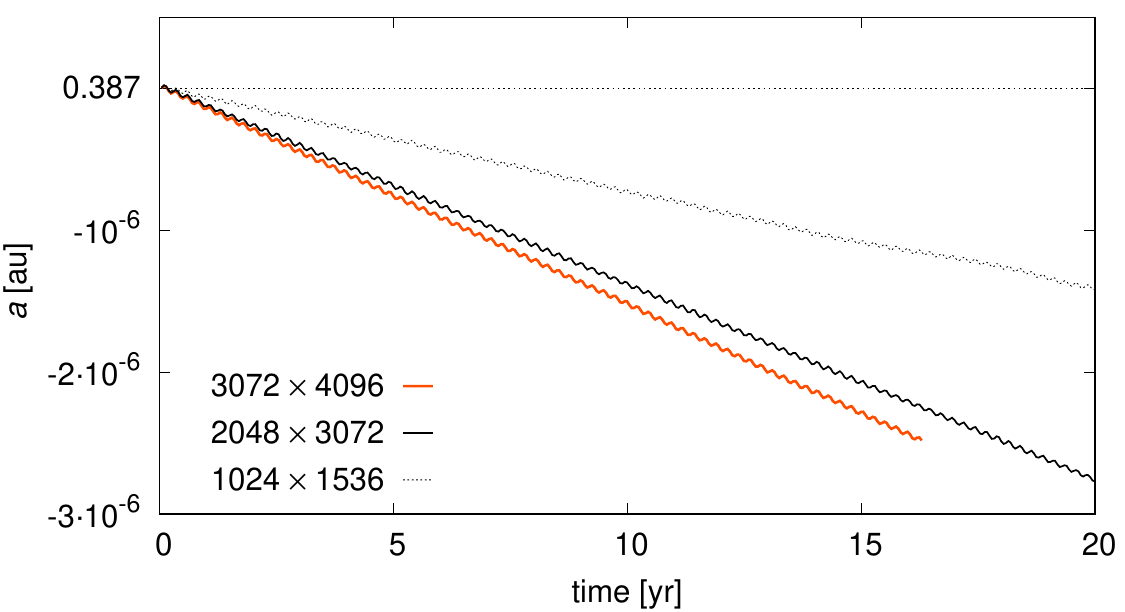} &
\includegraphics[width=8cm]{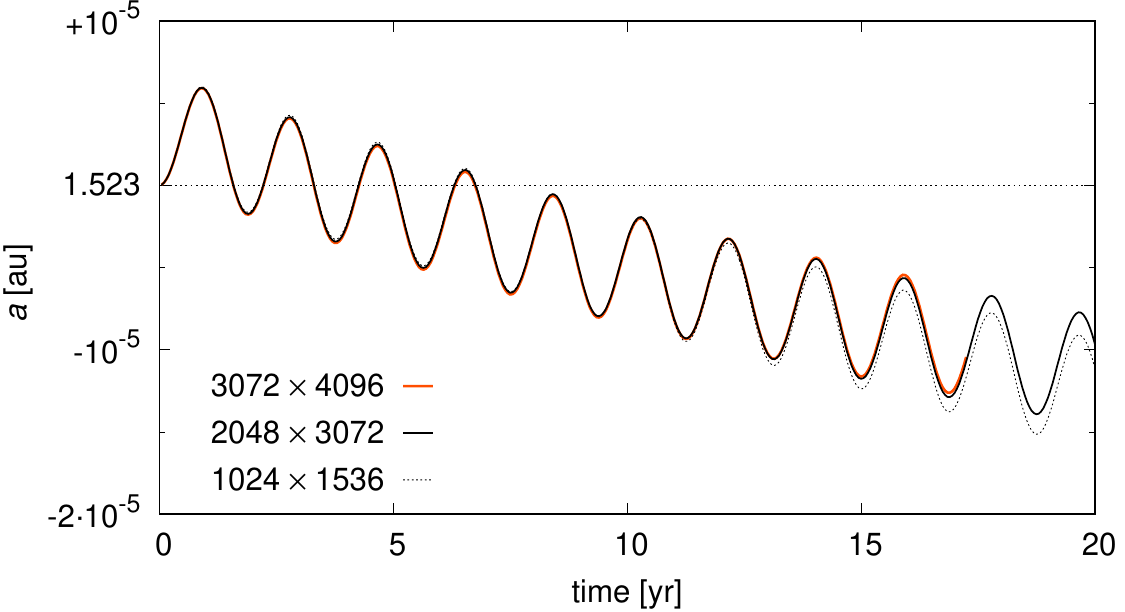} \\
\end{tabular}
\caption{Evolution of semimajor axis~$a$ (bottom)
and eccentricity~$e$ (top)
of a Mercury- (left)
and Mars-sized bodies (right) for three resolutions:
$1024\times 1536$ (low; dashed),
$2048\times 3072$ (medium; solid), and
$3072\times 4096$ (high; red).
In particular, there is a difference in the migration rate of Mercury,
which is about 2~times lower for the low-resolution simulation.
Nevertheless, it is always significantly smaller than that of Mars
(by an order of magnitude).
The short-term evolution of eccentricity exhibits the same amplitude
for all resolutions.
}
\label{test_mercury}
\end{figure}

For the smallest protoplanets of Mercury-size, the resolution of hydrodynamical
simulations is critical to correctly capture not only the Lindblad torque
(spiral arms) but also the corotation torque, which arises in a relatively
narrow range of radii. Alternatively, if we deliberately use a low resolution
for some simulation (or a part of it), we have to control and track the
respective discretisation error. Our convergence test is thus designed as follows:
we assume the same disk as in the main text,
spanning from 0.2 to 2.0\,au,
Mercury-size protoplanet at 0.387\,au,
three different resolutions:
$1024\times1536$ (low),
$2048\times3072$ (medium), and
$3072\times4096$ (high),
all with a logarithmic radial spacing.
This corresponds to approximately 3, 6, 10 cells per Hill sphere, respectively.
The results are summarized in Figure~\ref{test_mercury}.
It turns out that the migration rate of Mercury-size protoplanet
is underestimated by a factor of 2 in low-resolution simulations.

Fortunately, our interpretation of results does not depend sensitively
on the particular value of the migration rate of Mercury. It is rather
based on the fact that its migration rate (in absolute sense) is significantly
smaller that the rate for more massive protoplanets (Mars-size, or later Venus-size),
which is true both in low- and high-resolution simulations.

Apart from this convergence test for single planets,
we also performed similar tests with multiple planets
at higher-than-nominal resolution,
namely $2048\times3072$ (medium) instead of $1024\times1536$ (low),
for a relatively short time span only.
It turns out the evolution is qualitatively the same,
although interacting individual bodies may exhibit different
trajectories due to the chaos present in practically all
N-body systems.

%%%%%%%%%%%%%%%%%%%%%%%%%%%%%%%%%%%%%%%%%%%%%%%%%%%%%%%%%%%%%%%%%%%%%%%%

\section{Parameters of N-body simulations.}

Generally, parameters of the N-body simulations should correspond
to the hydrodynamic simulations above, but it is also possible
to vary them and perform an extended study (see Sec.~\ref{sec:additional}).
Here we summarize the nominal values.
The disk lifetime is assumed $t_2-t_1 = 10\,{\rm Myr}$;
migration time scale is approximated as $\tau(M)=0.2\,{\rm Myr}/(M/M_{\rm E})$,
migration rate $\dot a(a-r_0,\tau) = -{\rm sgn}(a-r_0)(a-r_0)^2/\tau$;
0-torque radius $r_0 = 1.1\,{\rm au}$ (fixed in this case),
however, for the eccentricity $e > 0.02$
we apply a reduction of the torque
($r_0 = 0\,{\rm au}$, $\tau = 2\,{\rm Myr}/(M/M_{\rm E})$),
and for $e > 0.06$ the Lindblad torque reversal
($\dot a \simeq 0$);
damping of eccentricity $\tau_{\rm e} = 1\,{\rm My}$,
which is only active for $e > e_{\rm hot} = 0.02$;
damping of inclination $\tau_{\rm i}$ is the same,
with $i_{\rm hot} = 0$.
The time step is $\Delta t = 0.01\,{\rm yr} \doteq 1/20 P_{\rm orb}$ at $0.387\,{\rm au}$,
with subdivisions during encounters.

The actual transversal, radial and vertical acceleration terms are computed as follows:
\begin{equation}
\ddot{\vec r_i} = {1\over 2} \dot a_i {G(M_\oplus+M_i)\over a_i^2 |\vec v_i|}\,\vec v_i
-2\kern1pt{\cal H}(e_i - e_{\rm hot}){1\over\tau_{\rm e}}{\vec r_i\cdot\vec v_i\over r_i^2}\,\vec r_i
-2\kern1pt{\cal H}(i_i - i_{\rm hot}){1\over\tau_{\rm i}}\vec v_i\cdot\hat z\hat z\,,
\end{equation}
where $a_i$ denotes the semimajor axis of the body~$i$,
$\vec r_i$~its position,
$\vec v_i$~velocity, and
${\cal H}$~the Heaviside step function.
Optionally, we apply an additional forcing of eccentricity and inclination,
simply using $2{\cal H}-1$ instead of ${\cal H}$.
In order to account for pebble accretion we
increase the mass of bodies as
$\dot M_i = f\dot M_{\rm p}$,
where $f$ denotes the filtering factor, reaching up to 5\,\%
and linearly depending on~$M_i$.
Because $f$ values are low, we assume the pebble flux
$\dot M_{\rm p}$ is practically independent of~$r$.

%%%%%%%%%%%%%%%%%%%%%%%%%%%%%%%%%%%%%%%%%%%%%%%%%%%%%%%%%%%%%%%%%%%%%%%%

\section{Additional simulations}\label{sec:additional}

\paragraph{Unstable disks with variable $\alpha$-viscosity.}
The kinematic viscosity~$\nu$, or $\alpha$, is another important parameter
which determines the overall structure of the disk, by means
of viscous heating. Apart from the (negligible) molecular viscosity,
it should account for (unresolved) turbulence and should be
thus considered as an eddy viscosity. Its value is not well constrained,
also because the underlying turbulence is not fully understood\cite{Nelson_etal_2013MNRAS.435.2610N,Klahr_Bodenheimer_2003ApJ...582..869K,Balbus_Hawley_1991ApJ...376..214B,Bae_etal_2016ApJ...833..126B}.

A key question then is whether $\alpha$ is constant, or rather temperature-dependent $\alpha(T)$.
If the turbulence in the inner disk is induced by the magneto-rotational instability (MRI)
which indeed depends on the ionisation (and $T$),
we should expect a viscosity `bump'\cite{Flock_etal_2016ApJ...827..144F}.
with a corresponding drop in the surface density.
We already tested a disk with variable~$\alpha$,
parametrized as in \cite{Flock_etal_2016ApJ...827..144F}
by $\alpha_{\rm in} = 1.9\times10^{-2}$,
$\alpha_{\rm out} = 10^{-3}$,
$\Delta T = 25\,{\rm K}$,
and $T_{\rm MRI} = 1000\,{\rm K}$,
and performed a 1D relaxation procedure,
but we realised the disk is unstable on the viscous time scale,
periodically changing the surface density~$\Sigma(r)$ profile
and thus protoplanet migration.
Such an instability is also present in similar 3D models (Flock, pers.~comm.).
This may have implications for the terrestrial zone, and
it seems worth of a separate study.

\paragraph{N-body simulations vs migration parameters.}
In order to improve the statistics, we performed N-body simulations
of low-mass protoplanets (${<}\,0.1\,M_{\rm E}$),
in which we changed one (or two) parameters. For example, we used 
Mercury-size protoplanets with $m = 0.05\,M_{\rm E}$;
migration time scale $\tau(m)=0.1\,{\rm My}/(m/M_{\rm E})$, or fixed at 0.1\,Myr, 1.0\,Myr;
migration rate $\dot a(a-r_0,\tau)$ linear, or 'sinusoidal';
0-torque radius as a range $0.7$ to $1.0\,{\rm au}$, $r_0(m)=\max(0.7+1.5\,(m/M_{\rm E}-0.8),0.0)\,{\rm au}$, square-root;
damping of eccentricity $\tau_e = 0.1\,{\rm My}$, $0.01$,
for $e > e_{\rm hot} = 0.04$, or $e(m)=0.01+0.03 (m/M_{\rm E}-0.1)$;
similarly damping of inclination $\tau_i$,
with $i_{\rm hot} = 0.01$.
On the other hand, we used only a single value of the random seed
which determines the initial conditions (angles).

\begin{figure}
\centering
\begin{tabular}{l}
\includegraphics[width=8cm]{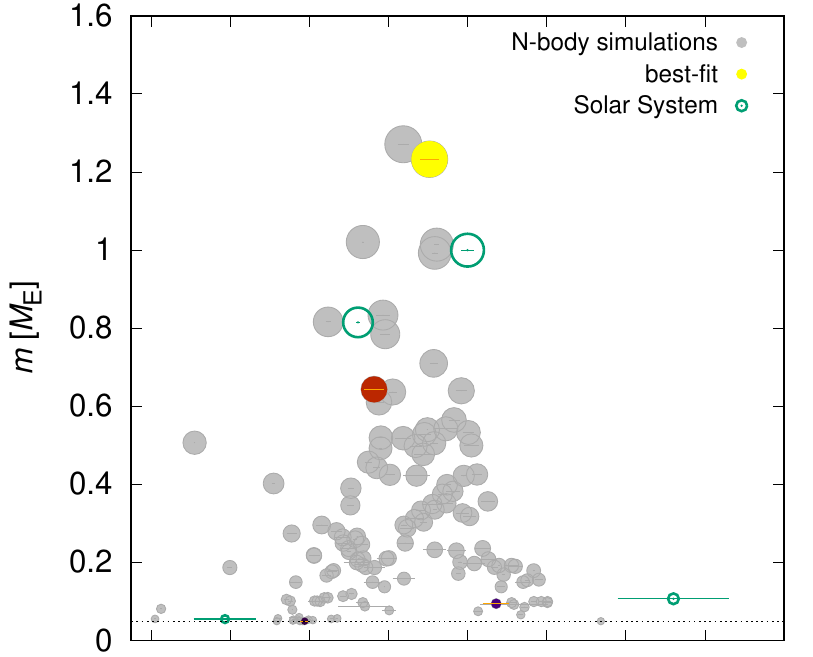} \\[-0.3cm]
\includegraphics[width=8cm]{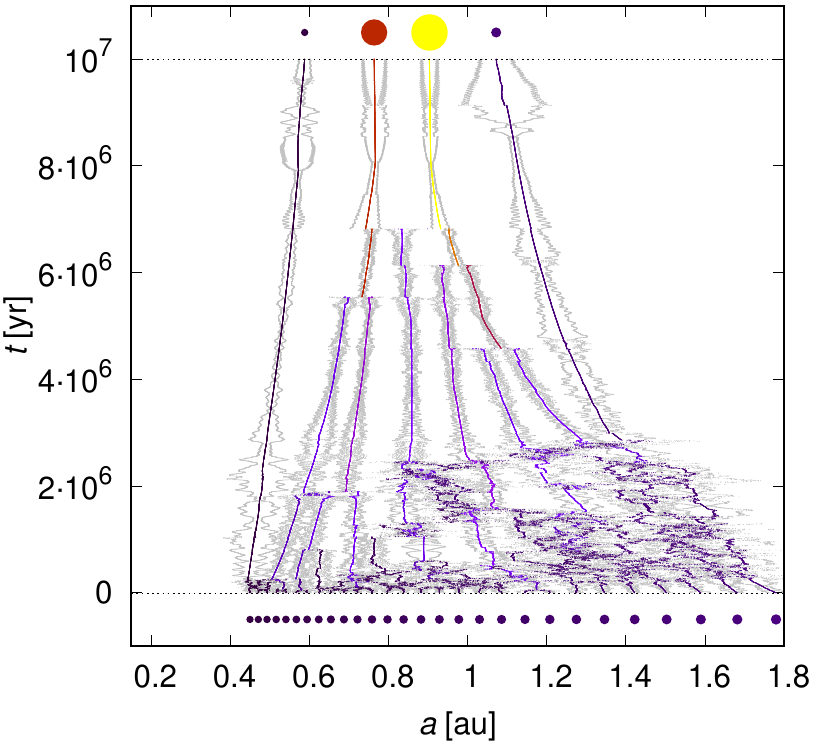}
\end{tabular}
\caption{
Final mass $m$ vs~the final semimajor axis $a$ (top)
and $a$ vs time~$t$ (bottom),
for an extended set of 20 N-body simulations with prescribed migration,
with different values of parameters (gray)
and one selected 'best-fit' simulation (colour).
The Solar System is shown for comparison (green).
}
\label{symba9_J}
\end{figure}

As expected, the outcomes of our simulations are more diverse,
because the parameters do affect the migration (Figure~\ref{symba9_J}).
Nevertheless, even this set often produces Venus and Earth analogues,
with Mercury and Mars analogues at the boundaries of the convergence zone.
In order to decide which parameters are more suitable for the Solar System,
it would be necessary to compute a subset of ${\sim}\,25$ simulations
with multiple random seeds for {\em every\/} parameter set,
since the evolution is chaotic --- especially due to close encounters,
which sensitively depend on the geometry --- and it is practically
impossible that every single simulation ends up as a perfect match.
Nevertheless, we find the overall evolution to be systematic
and capable of producing terrestrial planets with correct parameters.

\paragraph{Mercury scattering off Venus.}
The orbit of Mercury with its current osculating eccentricity $0.206$ and inclination $7.0^\circ$
poses a major observational constraint, because it cannot by possibly explained
by the hot-trail effect on Mercury which is too weak.
One needs a different mechanism and a close encounter with Venus (or a series of them)
is a logical choice.

\begin{figure}
\centering
\includegraphics[width=9cm]{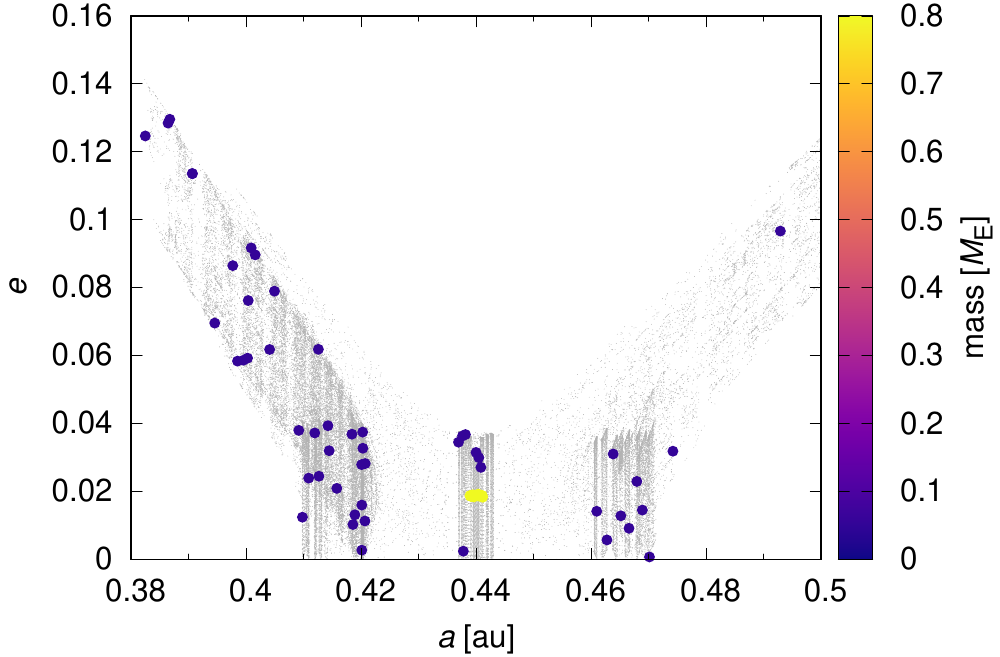}
\includegraphics[width=6.5cm]{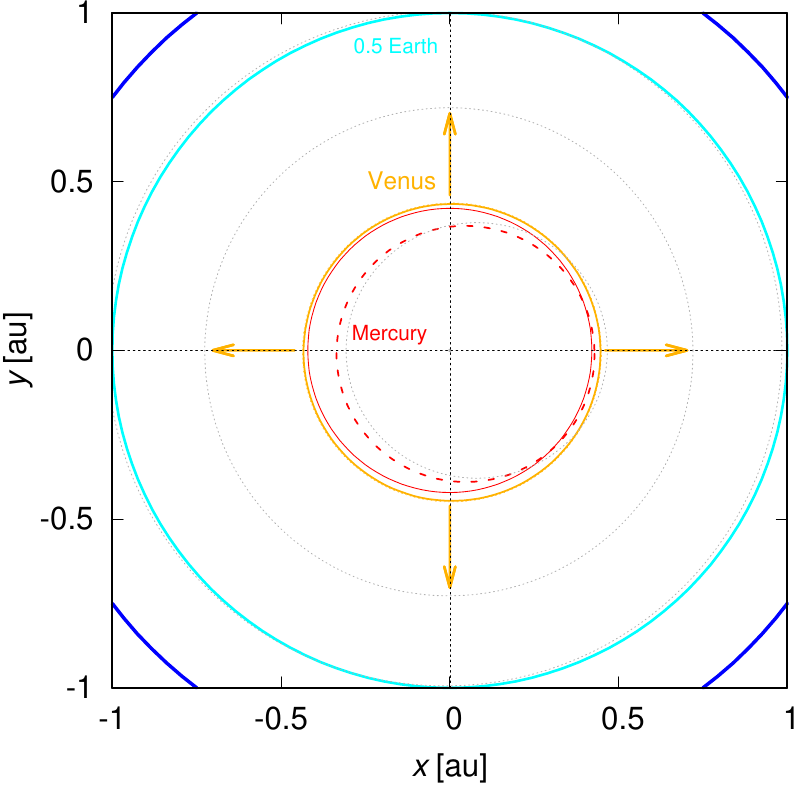}
\caption{Left: An overview of Mercury orbits $(a, e)$ which resulted from a series
of close encounters with Venus (shifted deliberately to $a_{\rm V} = 0.44\,{\rm au}$
and assumed to migrate outwards afterwards).
The initial position of Mercury was varied in the interval
$a_{\rm m} \in (0.41; 0.47)\,{\rm au}$.
The evolution over 10\,kyr was computed with a pure N-body model, without any hydrodynamics.
The final semimajor axis, eccentricity and inclination of Mercury may reach
$a_{\rm m}' = 0.387\,{\rm au}$, $e' \simeq 0.14$, $I' \simeq 2^\circ$.
A majority of bodies were scattered,
only a minority ended up as coorbitals or merged with Venus,
which is indicated by the colour scale (or mass).
Right: One example of Mercury scattering off Venus shown in the $(x, y)$ plane.
The orbits prior to the scattering (or rather a series of close encounters)
are plotted as solid lines; after the scattering as dashed lines.
The eccentricity of synthetic Mercury reaches $e' \simeq 0.14$.
For comparison, the proper eccentricity of observed Mercury is $e_{\rm prop} = 0.167$.
A further increase can be expected due to secular perturbations
and corresponding forced oscillations; the osculating $e_{\rm osc} = 0.206$.
The current orbits of planets are plotted in gray for comparison. 
The arrows indicate a subsequent migration of Venus
driven by the gas disk torques to its current position ($a = 0.723\,{\rm au}$).
}
\label{mercury_REBOUND_ae}
\end{figure}

To this point we performed a simplified scattering experiment ---
we moved Venus as close as $a_{\rm V} = 0.44\,{\rm au}$;
its $e = 0.02$, $I = 0.004\,{\rm rad}$ slightly increased due to the hot-trail,
acting both in the mid-plane and in the vertical direction\cite{Eklund_Masset_2017MNRAS.469..206E}.
We varied the initial position Mercury in the interval
$a_{\rm m} \in (0.41; 0.47)\,{\rm au}$.
We used the Rebound integrator\cite{Rein_Spiegel_2015MNRAS.446.1424R} alone,
i.e. N-body without any hydrodynamics
which is a good approximation for the scattering itself (although cf.~\cite{Broz_etal_2018A&A...620A.157B}; Fig.~6).
However, the integration time span was $10\,{\rm kyr}$ because the encounters are repeated.
Venus is much more massive than Mercury ($0.8$ vs $0.05\,M_{\rm E}$)
so its orbit does not change much,
but the potential well of Venus is just deep enough to increase Mercury's
$e_{\rm m}' \simeq 0.14$, $I_{\rm m}' \simeq 2^\circ$
(see Figure~\ref{mercury_REBOUND_ae}).

\begin{figure}
\centering
\includegraphics[width=9cm]{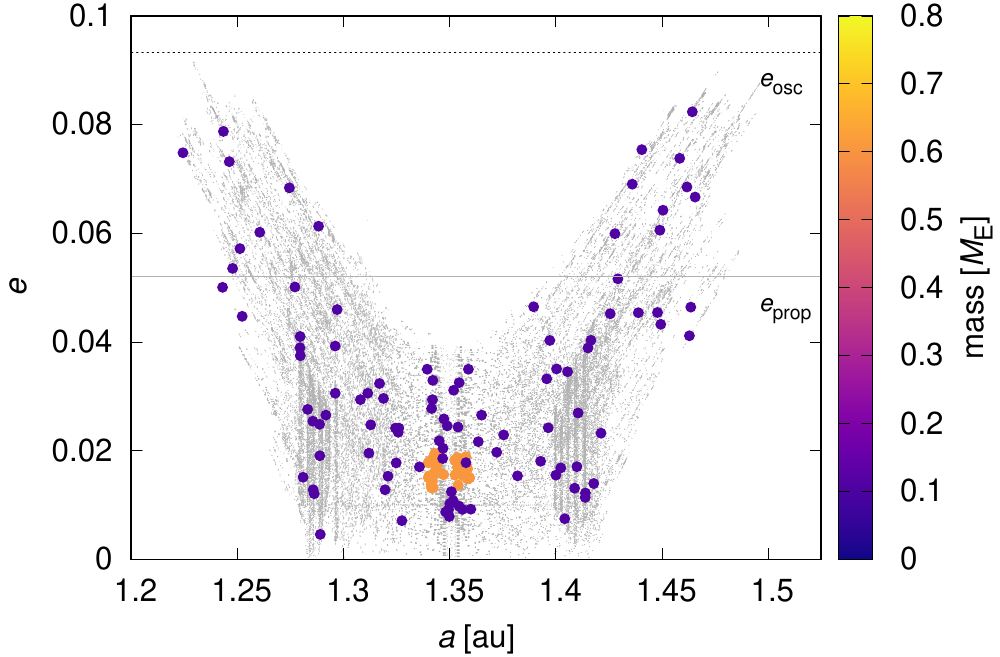}
\includegraphics[width=6.5cm]{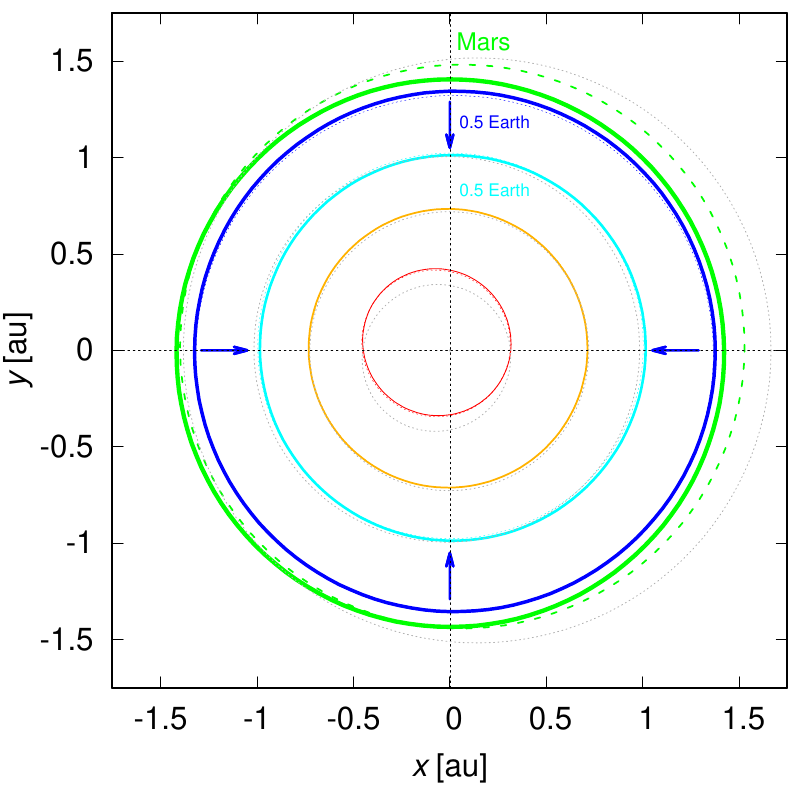}
\caption{The same as Fig.~\ref{mercury_REBOUND_ae}
but for Mars scattering by 0.5 Earth (shifted to $a_{\rm E} = 1.35\,{\rm au}$).
The proper eccentricity $e_{\rm prop} = 0.052$
and the osculating $e_{\rm osc} = 0.093$
of observed Mars are indicated by horizontal lines.}
\label{mars_REBOUND_ae}
\end{figure}

While this seems to be slightly lower than the observed osculating values,
$e_{\rm osc} = 0.206$, $I_{\rm osc} = 7.0^\circ$,
there are long-term secular perturbation induced by the whole planetary system
and our values are just at the lower limit of the respective oscillations,
$e \simeq 0.14$ to~$0.30$, $I \simeq 1^\circ$ to~$10^\circ$
(e.g. \cite{Laskar_1988A&A...198..341L}).
More rigorously, the comparison should be performed with respect to the proper elements,
$e_{\rm prop} = 0.167$, $I_{\rm prop} = 6.8^\circ$,
but the latter depends on the position of the invariable plane.

Finally, Venus and Mercury have to be {\em detached} from each other
otherwise they would still interact. Luckily, the migration of Venus is fast enough,
$\d a/\d t \simeq 10^{-6}\,{\rm au}\,{\rm yr}^{-1}$,
and in the correct direction (outward) in the MRI-active disk (Fig.~2 from the main text).
It would take only about $(0.723-0.440)/(\d a/\d t) \doteq 10^5\,{\rm yr}$
to migrate to its current position, $a_{\rm V} = 0.723\,{\rm au}$.
Consequently, the scattering by Venus is a very natural explanation
for the origin of Mercury's orbit. Alternatively, see \cite{Roig_etal_2016ApJ...820L..30R}.

Similar arguments are also valid for Mars (see Figure~\ref{mars_REBOUND_ae}).

\paragraph{Orbital elements changes by impacts.}

To verify  changes of the semimajor axis, eccentricity, and inclination
due to mutual collisions of protoplanets, we may use the Gauss equations:
\begin{eqnarray}
\Delta a &=& {2\over n\eta} [\Delta v_{\rm T} + e(\Delta v_{\rm T}\cos f + \Delta v_{\rm R}\sin f)]\,,\\
\Delta e &=& {\eta\over na} [\Delta v_{\rm R}\sin f + \Delta v_{\rm T}(\cos f+\cos E)]\,,\\
\Delta i &=& {1\over na\eta} \Delta v_{\rm W} {r\over a} \cos(\omega+f)\,,
\end{eqnarray}
where
$\Delta v_{\rm R}$,
$\Delta v_{\rm T}$,
$\Delta v_{\rm W}$
denote the radial, transversal and vertical components of the velocity difference,
$n$~mean motion,
$\eta \equiv \sqrt{1-e^2}$,
$f$~true anomaly,
$E$~eccentric anomaly and
$\omega$~argument of pericentre.

For a canonical Moon-forming impact with masses
$m_2 \simeq 0.1\,M_{\rm E}$,
$m_1 \simeq 1.0\,M_{\rm E}-m_2$,
$a = 1\,{\rm au}$,
$e = 0.04$,
and geometry $f = 0$, $E = 0$, $\omega = 0$,
we may estimate the velocity difference
$\Delta v \simeq (m_2/m_1) na(\sqrt{(1+e)/(1-e)}-1) = 0.13\,{\rm km}\,{\rm s}^{-1} = 0.01\,v_{\rm esc}$%
\footnote{although the true impact velocity is by $v_{\rm esc}$ larger due to gravitational focussing}
and, assuming the velocity components are of the same order, then
$\Delta a \simeq 0.009\,{\rm au}$,
$\Delta e \simeq 0.009$,
$\Delta i \simeq 0.004\,{\rm rad}$,
which can be considered small compared to changes induced by hydrodynamics
(Type-I migration, damping, the hot-trail effect).
On contrary, for an half-Earth impact\cite{Canup_etal_2019},
much larger changes can be expected,
$\Delta a \simeq 0.082\,{\rm au}$,
$\Delta e \simeq 0.078$,
$\Delta i \simeq 0.039$.
However, if they occur early in a gas disk,
they do not matter much,
because the final values are regulated by hydrodynamics anyway.

\paragraph{Depleted asteroid belt.}
Apart from the terrestrial convergence zone, there is another one beyond
the snowline in the giant-planet zone\cite{Chrenko_etal_2017A&A...606A.114C}.
The region between them, where the current main asteroid belt is located,
can be logically considered a {\em divergence\/} zone.\cite{Bitsch_etal_2014A&A...564A.135B}
To verify this point, we prepared another simulation with 20~Mars-size protoplanets,
spanning 1.5 to 4\,au with spacing of 5~mutual Hill radii,
and supplemented by 100 of $D = 100\,{\rm km}$ asteroids,
as representatives of the former main-belt population.
The disk is an extension of the nominal disk with the same $\Sigma_0$ as above.
We confirm that out of 20 protoplanets most of them migrate away, at a rate
$\d a/\d t \doteq 3\times 10^{-6}\,{\rm au}\,{\rm yr}^{-1}$,
and the respective zone is divergent.
Moreover, if the runaway growth occurs in both convergence zones,
the mass depletion in between can easily reach a factor of 100
(as in \cite{Broz_etal_2018A&A...620A.157B}; Fig.~20).

\begin{figure}
\centering
\begin{tabular}{l}
\includegraphics[width=9.65cm]{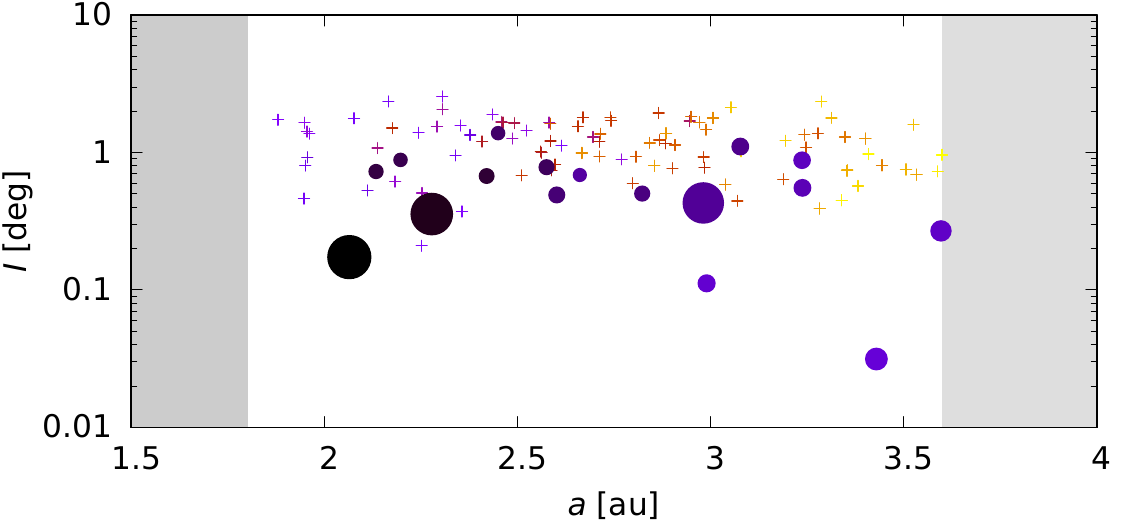} \\[-0.2cm]
\kern-.1cm\includegraphics[width=11cm]{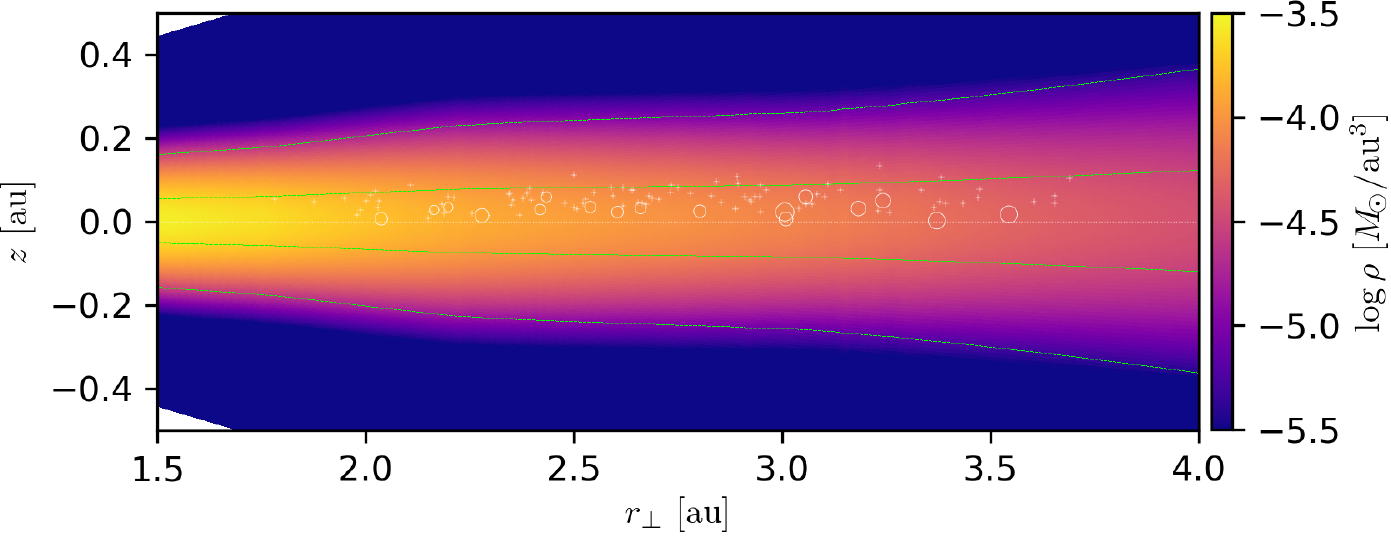}
\end{tabular}
\caption{
Asteroid belt region in which Mars-sized protoplanets and
$D = 100\,{\rm km}$ asteroids mutually interact,
shown in the semimajor axis~$a$ vs inclination~$I$ (top)
and the gas volumetric density $\rho(r_\perp,z)$ plots (bottom),
given by the exponential profile and the local scale height~$H$.
Initial conditions were with low~$I < 0.01^\circ$,
the situation corresponds to $t = 24\,{\rm kyr}$.
We assume the nominal disk with $\Sigma_0 = 750\,{\rm g}\,{\rm cm}^{-2}$,
only extended up to 4\,au.
According to our hydrodynamical model, the protoplanets migrate
at a rate of the order of $3\times 10^{-6}\,{\rm au}\,{\rm yr}^{-1}$,
and they slowly move away;
the asteroid belt is thus effectively a {\em divergence\/} zone.
On the way, they scatter the asteroids,
and push them to higher~$I \simeq 3^\circ$ (and also~$e \simeq 0.1$).
If we extrapolate this random-walk evolution up to $t = 300\,{\rm kyr}$,
assuming $\sqrt{t}$ dependence, it may reach the observed values
$I \simeq 10^\circ$ (and $e \simeq 0.3$).
%The snowline is located at 2.2\,au in this model.
}
\label{mainbelt}
\end{figure}

\paragraph{Inclined asteroid belt.}
On their way, protoplanets excite asteroids and
virialisation between high- and low-mass bodies,
close encounters,
and mean-motion resonances
push both $e$ and $I$ up to $0.1$ and $3^\circ$,
respectively, in the course of time (Figure~\ref{mainbelt}).
A damping of $e$ and $I$ due to density waves\cite{Tanaka_Ward_2004ApJ...602..388T}
is not efficient for asteroids (having a small factor $m/M_\odot$).
On the other hand, aerodynamic drag is not yet strong enough,
having a time scale of $\tau_{\rm drag} \simeq 1\,{\rm Myr}$ for $D = 100\,{\rm km}$ bodies.
The time span of our simulation ($24\,{\rm kyr}$) is again relatively short,
but if we extrapolate to $1\,{\rm au}/(\d a/\d t) \doteq 3\times10^5\,{\rm yr}$,
assuming a square-root dependence~$\sqrt{t}$ suitable for a random-walk process,
it would lead to the final $e \simeq 0.3$ and $I \simeq 10^\circ$,
in agreement with observations.
A single embryo remaining close to the (unstable) 0-torque radius,
may be eliminated later, when the disk structure changes (e.g. due to decreasing $\Sigma$),
or eventually by giant-planet migration which is needed to shape
the Kuiper belt\cite{Nesvorny_Vokrouhlicky_2016ApJ...825...94N}.

%%%%%%%%%%%%%%%%%%%%%%%%%%%%%%%%%%%%%%%%%%%%%%%%%%%%%%%%%%%%%%%%%%%%%%%%

%\fi
%\setcounter{section}{8}

\section{Geochemical constraints}

There are several lines of evidence that protoplanets formed early.
For example, $^{182}$Hf-$^{182}$W systematics applied to SNC meteorites indicates
Mars formed on a~time scale of only 2\,My\cite{Dauphas_Pourmand_2011Natur.473..489D}.
Ratios of $^{3}$He/$^{22}$Ne and $^{20}$Ne/$^{22}$Ne in Earth's mantle
can be explained by the dissolution of nebular gas in magma oceans
on Earth or precursor protoplanets.\cite{Tucker_Mukhopadhyay_2014E&PSL.393..254T,Wu_etal_2018JGRE..123.2691W}.
Similarly, the atmospheric ratios of $^{20}$Ne/$^{22}$Ne and $^{36}$Ar/$^{38}$Ar
of Venus indicate escape of a 100-bar primordial (H$_2$) atmosphere,
which can only be captured from the nebula if protoplanets were massive,
i.e. ${>}\,0.6\,M_{\rm E}$\cite{Lammer_etal_2018A&ARv..26....2L}.

\begin{figure}
\centering
\includegraphics[width=10cm]{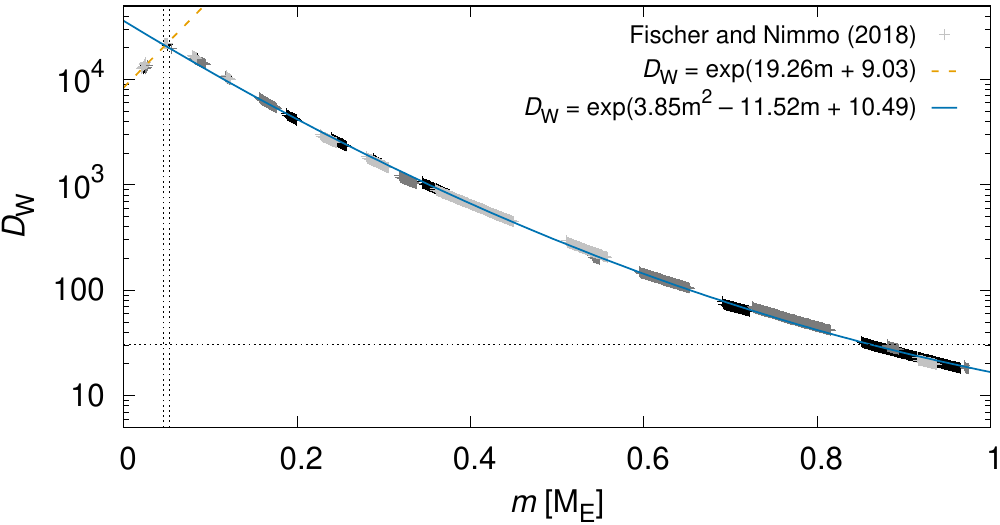}
\caption{
Variable partitioning coefficient of tungsten~$D_{\rm W}$
as a function of mass~$m$ of the growing Earth.
Data (crosses) from ref.~\cite{Fischer_Nimmo_2018E&PSL.499..257F};
fits (lines) correspond to Eq.~(\ref{eq:D_W}).
}
\label{DW_fit}
\end{figure}

Our simulation predict the rapid growth of planets, which can be punctuated by
a late giant impact long after dissipation of nebular gas.
We used the geochemical model for the Hf/W system\cite{Yu_Jacobsen_2011PNAS..10817604Y,Jacobsen_2005AREPS..33..531J}
to evaluate whether our N-body accretional simulations
are consistent with the observed anomaly of radiogenic tungsten $^{182}$W,
$\varepsilon_{\rm 182W} \equiv \left\{{[^{182}{\rm W}/^{184}{\rm W}] \over [^{182}{\rm W}/^{184}{\rm W}]_{\rm CHUR}} -1\right\}\cdot 10^4 = 1.9\pm0.1\,$\cite{Kleine_etal_2002Natur.418..952K},
where CHUR denotes the chondritic uniform reservoir.
We used the accretion histories of Earth analogs obtained in our {\it SyMBA} simulations,
and assumed either a~constant silicate/metal partitioning coefficient,
$D_{\rm W} = 31$, or a~variable~$D_{\rm W}$, which depends on the
mass~$m$ of the growing Earth according to the formula:
\begin{equation}
D_{\rm W}(m) = {\rm min}(\exp(19.26\,m + 9.03), \exp(3.85\,m^2 - 11.52\,m + 10.49))\,,\label{eq:D_W}
\end{equation}
obtained from data sent to us by R.~Fischer\cite{Fischer_Nimmo_2018E&PSL.499..257F}
that correspond to the 3 simulations depicted in their Fig.~1 (see Figure~\ref{DW_fit}).

%Alternatively, $D_{\rm W}$ depends on the pressure~$P_{\rm CMB}$ at the core/mantle boundary,
%the temperature~$T_{\rm CMB}$,
%as well as the oxygen fugacity~$f_{\rm O2}$.
%The iron concentration in matle is assumed to be
%0.0101 prior to Moon-forming impact
%and 0.0626 afterwards\cite{Yu_Jacobsen_2011PNAS..10817604Y};
%we dot use the 0.14 value for $m < 0.2\,M_{\rm E}$,
%because a sudden change of~$f_{\rm O2}$ is considered unlikely.
%The activity coeffcient $\gamma_{\rm FeO} = 3$.
%After the Moon-forming impact,
%the equlibration proceeds at a reduced CMB pressure of 40\,GPa\cite{Yu_Jacobsen_2011PNAS..10817604Y}.
%For simplicity, we assumed full equilibration
%of (emulsified) impactor core with Earth's mantle.

\begin{figure}
\centering
\includegraphics[width=14cm]{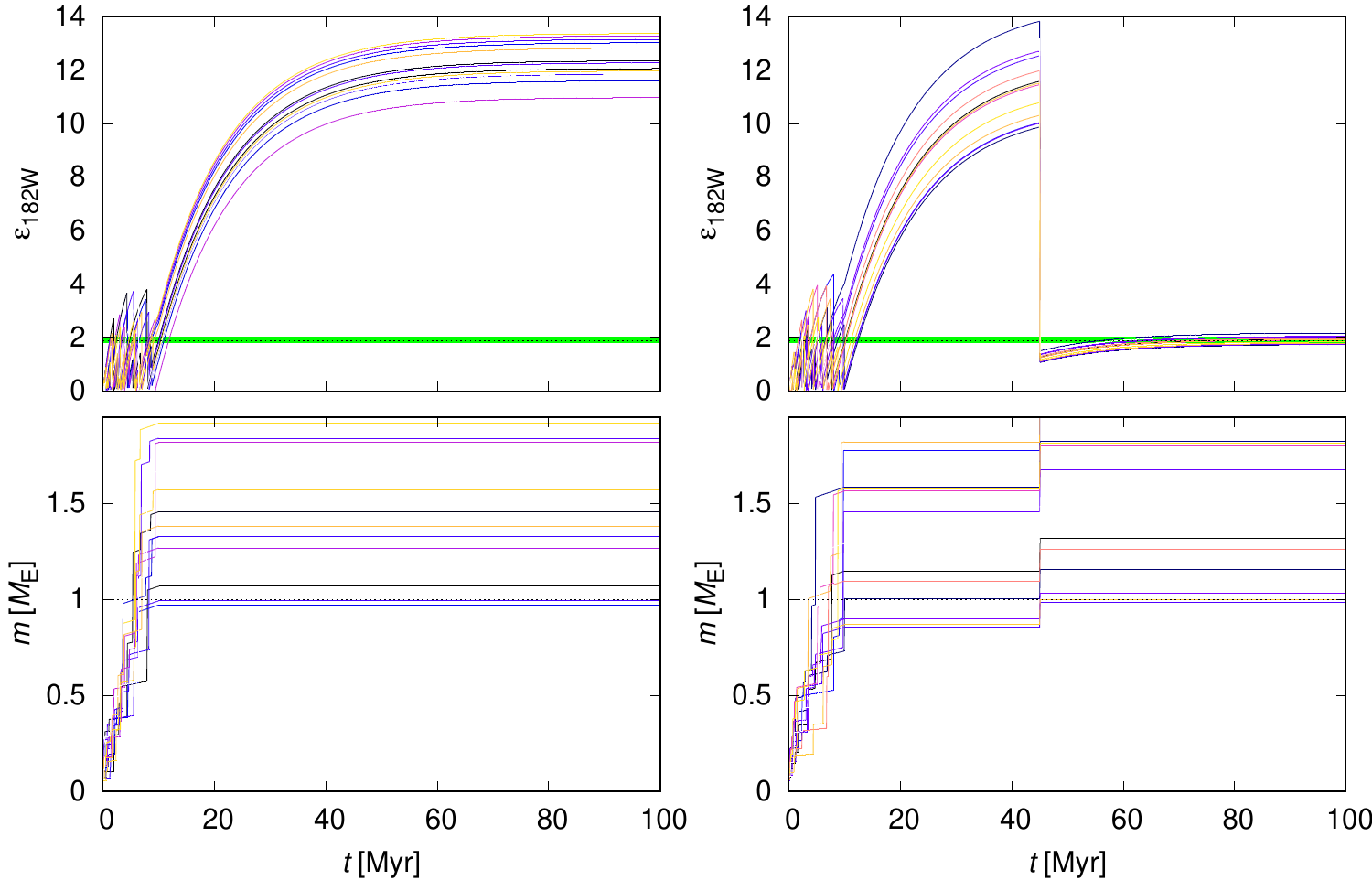}
\caption{
Accretion history $m(t)$ (bottom)
and tungsten anomaly $\varepsilon_{\rm 182W}$ (top)
for a subset of N-body simulations ending with 4~(left)
or 5~planets (right).
Colours correspond to individual simulations.
The observed value of $\varepsilon_{\rm 182W}$ and its uncertainty
is plotted as a green strip. In this test,
the partitioning coefficient of tungsten~$D_{\rm W}$ was assumed
to be constant and the relative impactor size is~${\sim}\,0.15$.
}
\label{w182}
\end{figure}

\begin{figure}
\centering
\includegraphics[width=14cm]{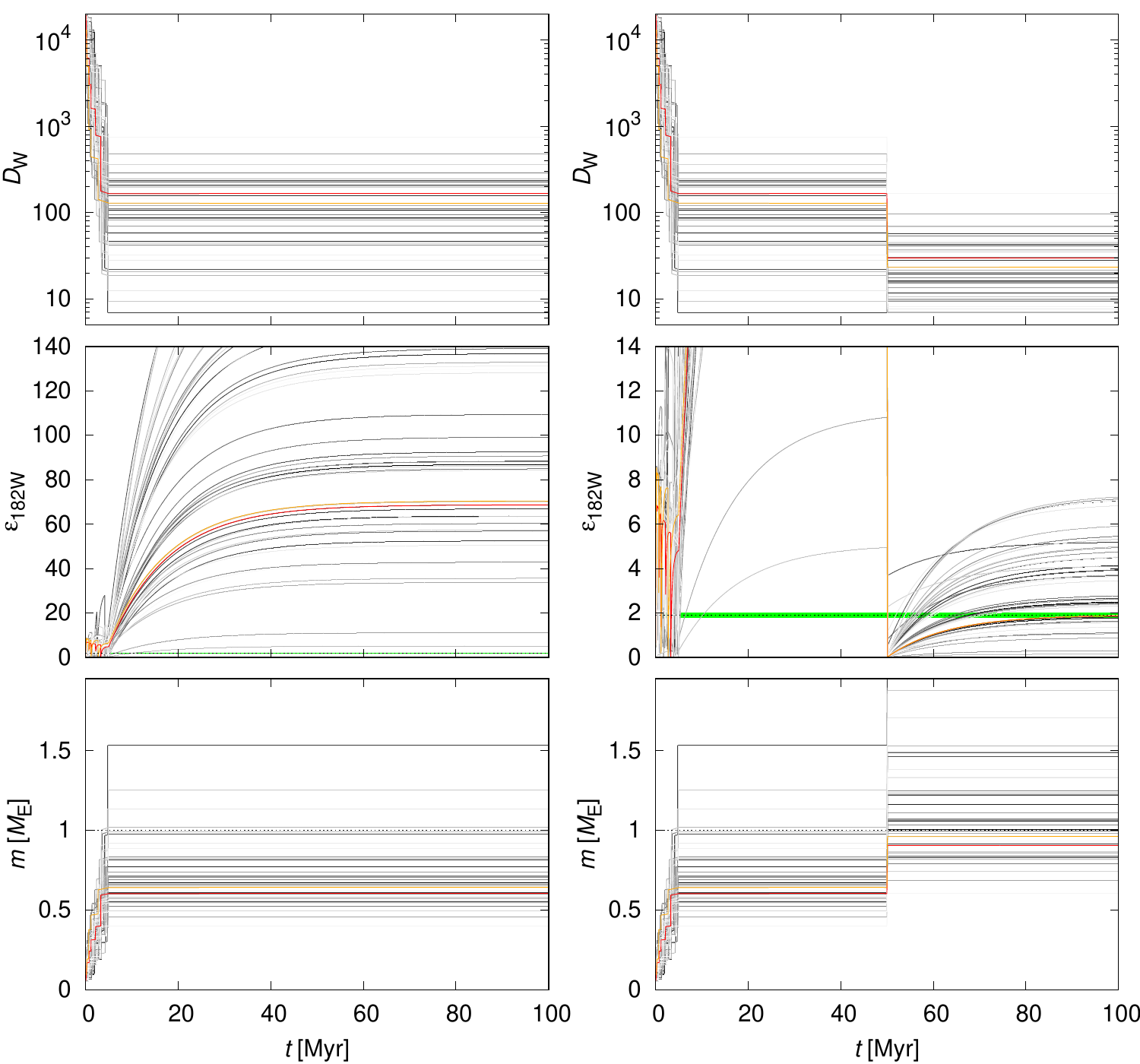}
\caption{
The same as Fig.~\ref{w182}, but with the partitioning coefficient~$D_{\rm W}$
as a function of planet mass~$m$\cite{Fischer_Nimmo_2018E&PSL.499..257F},
and the relative late impactor size~${\sim}\,0.33$.
}
\label{symba28c_pebbleflux2e-7_5MY_0.33SIZE_w182}
\end{figure}

Our results for 4- and 5-planet systems are shown in Figure~\ref{w182}.
The cases with 4~planets and no late impacts (${\sim}\,30\,\%$ of the simulations)
exhibit much higher final anomalies (12--14) than measured.
For the cases with 5~planets (another ${\sim}\,30\,\%$), we continued our N-body 
simulations until the (random) final impact and found a very wide range of 
$\varepsilon_{\rm 182W}$ values (as in \cite{Fischer_Nimmo_2018E&PSL.499..257F}; Fig.~5 therein).
We therefore preferred, in order to have better control of $\varepsilon_{\rm 182W}$,
to prescribe a late Moon-forming impact at $t_{\rm Moon}$.
The best results were obtained with $t_{\rm Moon}\simeq45$ Myr, where
practically all results are compatible with observations (Figure~\ref{w182}, right column).
This is close to a recent estimate of the time of Moon formation
${\simeq}\,50\,{\rm Myr}$ inferred from Hf-W systematics of lunar rocks
(\cite{Thiemens_etal_2019};
see however \cite{Touboul_etal_2015Natur.520..530T,Kruijer_etal_2015}).
For variable~$D_{\rm W}(m)$, there would be even higher anomaly,
because initially $D_{\rm W} \gg 31$,
the fractionation $f^{\rm Hf/W} \gg 15$,
and the derivative is
$\dot\varepsilon_{\rm 182W} \propto Q^\star_{\rm W} f^{\rm Hf/W}$
(\cite{Yu_Jacobsen_2011PNAS..10817604Y}; see Figure~\ref{symba28c_pebbleflux2e-7_5MY_0.33SIZE_w182});
this would require the relative mass ratio of impactors
0.33 to half-Earths\cite{Canup_etal_2019},
which are quite common in our simulations with convergent migration
(as already demonstrated in Figure~\ref{hist2_symba11b_reductions5}).
Earlier impacts would also require more massive impactors,
but there is a lower limit at approximately 30\,Myr (for half-Earths)
and an upper limit at 60\,Myr, provided that the effects of the late veneer
were minor. Indeed, the highly siderophile element (HSE) content in Earth's mantle
suggests the late veneer mass up to $0.005\,M_{\rm E}$\cite{Dauphas_Marty_2002JGRE..107.5129D,Dauphas_2017Natur.541..521D}
and accounting for this changes $\varepsilon_{\rm 182W}$ by only~${<}\,0.1$.

%\bibliography{references}
%\end{document}

%%%%%%%%%%%%%%%%%%%%%%%%%%%%%%%%%%%%%%%%%%%%%%%%%%%%%%%%%%%%%%%%%%%%%%%%

\section{Additional implications}

\paragraph{Small Mars.}
For Mars, there is an additional circumstance,
which contributes to its small size
(apart from being at the edge of the convergence zone).
Normally, pebble accretion proceeds for each planet in parallel
and the surface density of pebbles~$\Sigma_{\rm p}$ decreases only by a few per cent,
according to our estimate of the filtering factors (0.5~to 8\,\% for Mercury- to Venus-size).
But if pebbles are formed {\em only\/} within the terrestrial region (as a whole),
or when the flux of pebbles~$\dot M_{\rm p}$ becomes limited (e.g. by the gas giant core),
or exhausted in the course of time, the amount of inward-drifting pebbles
beyond the orbit is Mars is low, and a logical outcome could be small Mars.

\paragraph{Carbon corrosion.}
According to Zhu or Semenov opacities\cite{Zhu_etal_2012ApJ...746..110Z,Semenov_etal_2003A&A...410..611S},
there is an important transition due to carbon corrosion (or refractory organics).
Pebbles remain carbon rich until they reach about 1\,au, depending on the disk profile.
These pebbles are constantly reprocessed, because the pebble size
is fragmentation limited\cite{Birnstiel_etal_2012A&A...539A.148B},
so even carbon incorporated into pebbles can be eliminated
on the time scale shorter than drift.
Although we do not track the chemical composition explicitly,
only implicitly in terms of $\kappa$, an outcome could be
the carbon-poor composition of Earth\cite{Bergin_etal_2015PNAS..112.8965B}.

\paragraph{Dry Venus.}
A standard explanation for the non-existence of water vapour on Venus
is the runaway greenhouse effect,
followed by a dissociation of ${\rm H}_2{\rm O}$,
an escape of hydrogen, and
a drag-off of oxygen atoms\cite{Chassefiere_1997Icar..126..229C},
needed to explain the low abundance of oxygen in the atmosphere.
Alternatively, if the snowline was always located above 0.7\,au,
Venus naturally could not accrete icy pebbles and remained dry.
The runaway greenhouse effect in its atmosphere then would not
involve water vapour.

\paragraph{Dry inner belt.}
Pebble accretion is also inefficient for $D = 100\,{\rm km}$ asteroids
and when icy pebbles drift around, a vast majority is not captured.
Even Ceres with its mass $1.5\times10^{-4}\,M_{\rm E}$
is well within the Bondi regime of accretion;
the transition to the Hill regime occurs at approximately $10^{-2}\,M_{\rm E}$.
Moreover, by that time all asteroids are excited to high inclinations
$I \gg I_{\rm p} \doteq 0.005\,{\rm rad} = 0.29^\circ$
and orbit above/below the pebble disk (Fig.~\ref{mainbelt}),
which drastically reduces the efficiency of accretion.
The inner-main-belt asteroids thus could remain dry too.

%%%%%%%%%%%%%%%%%%%%%%%%%%%%%%%%%%%%%%%%%%%%%%%%%%%%%%%%%%%%%%%%%%%%%%%%

\let\oldthebibliography=\thebibliography
\let\oldendthebibliography=\endthebibliography
\renewenvironment{thebibliography}[1]{%
  \oldthebibliography{#1}%
}{\oldendthebibliography}

\bibliography{references}

\end{document}